\documentstyle[12pt,cite,epsfig]{article}
\long\def\Msun{~$M_{\odot}$}

\begin{document}


\begin{center}
{\large \bf BL~Lac Objects in the Synchrotron Proton
Blazar Model}\\[1cm] A. M\"ucke$^{a,}$\footnote{present
address: Ruhr-Universit\"at Bochum, Institut f\"ur Theoretische Physik, Lehrstuhl IV:
Weltraum- und Astrophysik, D-44780 Bochum, Germany\\
email: {\it{afm@tp4.ruhr-uni-bochum.de}}},
R.J. Protheroe$^b$, R. Engel$^c$, J.P. Rachen$^{d,}$\footnote{present address: T-Systems GEI GmbH, Munich, Germany\\
email: {\it{Joerg.Rachen@t-systems.com}}} \& T. Stanev$^c$\\
$^a$D\'epartement de Physique\\
Universit\'e de Montreal, Montreal, QC, H3C 3J7, Canada\\
$^b$Department of Physics and Mathematical Physics\\
The University of Adelaide, Adelaide, SA 5005, Australia\\
$^c$Bartol Research Institute, University of Delaware, Newark, DE 19716, USA\\
$^d$Sterrenkundig Instituut, Universiteit Utrecht, 3508 TA Utrecht, The Netherlands\\
\end{center}

\begin{abstract}
We calculate the spectral energy distribution (SED) of
electromagnetic radiation and the spectrum of high energy
neutrinos from BL Lac objects in the context of the Synchrotron
Proton Blazar Model.  In this model, the high energy hump of the
SED is due to accelerated protons, while most of the low energy
hump is due to synchrotron radiation by co-accelerated electrons.
To accelerate protons to sufficiently high energies to produce
the high energy hump, rather high magnetic fields are required.
Assuming reasonable emission region volumes and Doppler factors,
we then find that in low-frequency peaked BL~Lacs (LBLs), which
have higher luminosities than high-frequency peaked BL~Lacs
(HBLs), there is a significant contribution to the high frequency
hump of the SED from pion photoproduction and subsequent
cascading, including synchrotron radiation by muons.  In
contrast, in HBLs we find that the high frequency hump of the SED
is dominated by proton synchrotron radiation.  We are able to
model the SED of typical LBLs and HBLs, and to model the famous
1997 flare of Markarian 501.  We also calculate the expected
neutrino output of typical BL Lac objects, and estimate the
diffuse neutrino intensity due to all BL Lacs.  Because pion
photoproduction is inefficient in HBLs, as protons lose energy
predominantly by synchrotron radiation, the contribution of LBLs
dominates the diffuse neutrino intensity.  We suggest that nearby
LBLs may well be observable with future high-sensitivity TeV
gamma-ray telescopes.
\end{abstract}

\hspace*{2cm} Gamma-rays: theory, Neutrinos, Synchrotron emission, Cascade simulation

\date{Submitted to Astroparticle Physics, 15 March 2002}


\newpage

\section{Introduction}

 Blazars are identified as Optically Violent Variable (OVV)
 quasars (a sub-class of Flat Spectrum Radio Quasars, FSRQ) and
 BL Lacs which may be low-frequency or high-frequency peaked
 BL~Lac objects.  Their broad-band spectra consist of two
 spectral components which appear as broad `humps' in the
 spectral energy distribution, and are due to emission from a jet
 oriented at small angle with respect to the line-of-sight.  The
 low-energy component is generally believed to be synchrotron
 emission from relativistic electrons, and extends from the radio
 to UV or X-ray frequencies.  The origin of the high-energy
 component, from X-ray to $\gamma$-ray energies, in some cases 
 to several TeV, is still under debate.

 Various models have been proposed, with the most popular ones
 being the ``leptonic models'', where a relativistic
 electron-positron jet plasma up-scatters low energy photons to
 high energies via the Inverse Compton effect.  The emission
 region moves with relativistic velocities along the jet, and the
 resulting radiation is highly beamed into the observer's
 line-of-sight.  The target photon field could come either from
 the accretion disk surrounding the putative black hole (e.g.\cite{DS93,BMS97}),
 possibly partly reprocessed by broad-line region (BLR) clouds (e.g.\cite{sikora94})
 or a dusty torus (e.g. \cite{blaze2000}, \cite{DoneaProtheroe2002}), or it could be produced by the
 relativistic $e^\pm$ population itself, in so-called synchrotron
 - self Compton (SSC) models \cite{MGC92,BM96}. In such leptonic models, it seems
 plausible to assume that SSC radiation dominates in objects with
 relatively weak accretion disk radiation such as BL~Lac objects,
 while in FSRQs external photons provide the dominant target
 field for the Inverse Compton process.

 As an alternative to leptonic models, ``hadronic models'' were
 proposed more than 10 years ago to explain the $\gamma$-ray
 emission from blazars~\cite{MB89,M93}. Recently M\"ucke \& Protheroe
 \cite{MP2000,MP2001a} have discussed in detail the various
 contributing emission processes.  In hadronic models the
 relativistic jet consists of relativistic proton and electron
 components, which again move relativistically along the
 jet. High-energy radiation is produced through photomeson
 production, and through proton and muon synchrotron radiation,
 and subsequent synchrotron-pair cascading in the highly
 magnetized environment.  Again either external (i.e. from an
 accretion disk and/or IR-torus \cite{P96}) or internal photon
 fields (i.e. produced by synchrotron radiation from the
 co-accelerated electrons) can serve as the target for photopion
 production.  Gamma-ray loud BL Lac objects are most likely
 explained by the latter possibility. These models can, in
 principle, be distinguished from the leptonic models by the
 observation of high energy neutrinos generated in decay chains
 of mesons created in the photoproduction interactions (for a
 recent review see \cite{LM2000}).

 In this paper, we study the properties of the Synchrotron Proton
 Blazar (SPB) model \cite{MP2001a}, where the dominant target
 photon field is produced by directly accelerated electrons that
 manifests itself in the blazar SED as the synchrotron hump.  A
 detailed description of the model itself and its implementation
 as a Monte-Carlo/numerical code has already been given in
 \cite{MP2001a}.  Since this model is for objects with a
 negligible external component of the radiation field in the jet,
 we apply it only to BL~Lac objects.  It has been successfully
 demonstrated that this model reproduces well the observed
 double-humped blazar SED.  The goal of this paper is to present
 a comprehensive study of the model's parameter space, a
 promising tool for discriminating between leptonic models and
 the hadronic SPB-model.  We apply our model to both LBLs and
 HBLs, and discuss our results in the light of the suggested
 LBL/HBL continuity.
 
 In Section 2 we give a brief description of the SPB model. The model
 is applied to HBLs and LBLs to calculate the
 SEDs in Section 3. We vary the magnetic field strength and the
 target photon density and study their effect on the resulting
 cascade spectrum. These results are used to identify the
 parameter sets within the SPB-model which are typical for HBLs
 and LBLs in Sect. 3.2.  In Sect. 3.3 we compare the model
 predictions to observed SEDs from HBLs and LBLs.  One of the
 most dramatic properties of blazars is their variability, and
 this issue is addressed by modeling the evolution of the SEDs
 during outburst and quiescent stages in Sect. 3.4.  The
 predicted neutrino emission from these sources is calculated in
 Section 4. Finally, we discuss our results in Sect. 5.


\section{The model}

 We consider an emission region, or ``blob'', in an AGN jet which
 moves relativistically along the jet axis which is closely
 aligned with our line-of-sight. The model assumes that electrons
 ($e^{-}$) and protons ($p$) are co-accelerated at the same site
 in the jet. Due to pitch-angle scattering the resulting particle
 distributions are expected to be quasi-isotropic.  The energetic
 protons, which follow a power law energy spectrum, are injected
 instantaneously into a highly magnetized environment, and suffer
 from energy losses due to proton--photon interactions (meson
 production and Bethe-Heitler pair production), synchrotron
 radiation and adiabatic expansion.  The mesons produced in
 photomeson interactions always decay in astrophysical
 environments.
 For the magnetic fields and proton energies typically assumed in
 hadronic blazar models, however, secondary particles such as 
 mesons and muons may suffer synchrotron losses
 before they decay \cite{RM98}, and this is also taken into account.  The
 relativistic co-accelerated $e^{-}$ radiate synchrotron photons
 that serve as the target radiation field for proton-photon
 interactions and the pair-synchrotron cascade which subsequently
 develops.  This cascade redistributes the photon power to lower
 energies where the photons eventually escape from the emission
 region. The cascades can be initiated by photons from
 $\pi^0$-decay (``$\pi^0$ cascade''), electrons from the
 $\pi^\pm\to \mu^\pm\to e^\pm$ decay (``$\pi^\pm$ cascade''),
 $p$-synchrotron photons (``$p$-synchrotron cascade''), charged
 $\mu$-, $\pi$- \cite{RM98} and $K$-synchrotron photons
 (``$\mu^\pm$-synchrotron cascade'') and $e^\pm$ from the
 proton-photon Bethe-Heitler pair production (``Bethe-Heitler
 cascade''). The probability of pair production is calculated
 from the $\gamma\gamma$ pair production opacity.  The
 $e^+e^-$-pairs generated radiate synchrotron photons, which
 again suffer from pair production, and feed the cascade
 development.  We use the matrix method (e.g. \cite{PJ96}) for
 simulating the developing cascades. Our model utilizes exact
 cross sections pre-calculated using the Monte-Carlo technique.
 This is especially important for the hadronic particle
 production.  The photomeson production is simulated with the
 SOPHIA Monte-Carlo code \cite{SOPHIA}. Details of the model
 implementation, e.g. energy loss rates, cascading method, etc.,
 are described in \cite{MP2001a}.

 The ``$\pi^0$ cascades'' and ``$\pi^\pm$ cascades'' generate
 rather featureless photon spectra in contrast to
 ``$p$-synchrotron cascades'' and ``$\mu^\pm$-synchrotron
 cascades'' that produce a double-humped SED as typically
 observed for $\gamma$-ray blazars.  We find the contribution
 from Bethe-Heitler pair production to be negligible.  Direct
 proton and muon synchrotron radiation is mainly responsible for
 the high energy hump whereas the low energy hump is dominated
 by synchrotron radiation by the directly accelerated $e^-$, with
 a contribution of synchrotron radiation from secondary electrons
 (produced by the $p$- and $\mu^\pm$-synchrotron cascade).

 In the present paper, we have made three changes to the original
 model.  The first change addresses $e^+e^-$ pair production in
 photon-photon collisions. Due to energy and momentum
 conservation, one of the produced pair electrons (with energy
 $E_{e^+}$, $E_{e^-}$) carries most of the available energy.
 We take this into account by
 using $E_{e^+}=0.1E_{\gamma}$, whereas the original model
 assumed that $E_{e^+}=E_{e^-}=0.5E_{\gamma}$.  Test runs
 show that the effect of this improvement on the cascade
 development is very modest.

 The second change in the model concerns the particle acceleration
 time scale. In the most general form it can be written as
\begin{equation}
t'_{\rm{acc}} = \frac{\gamma' m c^2}{\eta e c^2 B'}
\end{equation}
 for almost all proposed particle acceleration scenarios, with
 $\eta\leq 1$ being the acceleration efficiency which is strongly
 model dependent and, in most cases, even not well defined.
 Hence, instead of specifying $\eta$, and thereby limiting
 ourselves to one specific scenario, we use Eq.~(1), and leave
 $\eta$ as a free parameter. The efficiency, $\eta$, essentially
 influences the cutoff energy of the accelerated particle
 spectrum which is obtained by balancing the loss and gain time
 scales.
 The particle spectrum is assumed to follow a Heaviside function at
 the cutoff energy.

 The third change concerns the treatment of the size of the
 emission region $R'$. In this paper we use $R'$ as a free
 parameter, independent of the variability time scale.  The
 justification for this is given in
 \cite{Protheroe_variabilitysize} which discuss the relation
 between the emission region geometry and the observed
 variability time.  Note that in blazar models where protons are
 picked up from the ambient medium by a relativistic blast wave,
 the so-called ``pick-up models'' \cite{schlicki}, the
 variability time scale is determined by density
 inhomogeneities in the ambient medium rather by the size of the
 emission region. The value of $R'$ is relevant for the
 $\gamma\gamma$ pair production opacity, $p\gamma$ interaction rate,
 $n\gamma$ opacity, adiabatic losses due to
 jet expansion (see Appendix A) and the normalization of the
 emitted cascade spectrum.

 Because of the high energy threshold for photoproduction,
 hadronic models require extremely high proton energies which can
 only be achieved in a highly magnetized environment, and so
 synchrotron losses become very severe. Magnetic field values of
 order $10^4$~G are expected near the horizon of a supermassive
 black hole with mass $10^8-10^9$\Msun \cite{BZ77}. Assuming
 magnetic flux conservation, jet magnetic fields may reach values
 of 1-100 G in an emission region $\simeq 100-1000$ AU away from
 the black hole horizon.

 Throughout this paper we assume that the injected and
 accelerated particle spectrum can be represented as a power-law
 spectrum $n'_p \propto \gamma_p^{'-\alpha_p}$,
 ${\gamma'_p}_1 \leq \gamma'_p \leq {\gamma'_p}_2$ and we use
 $\alpha_p=2$ and $\eta=1$ unless noted otherwise.
 In the following all quantities in the jet frame are indicated by a
 prime, while quantities in the observer's frame are unprimed.


\section{Applications}

\subsection{Observations of HBL and LBL}

 Gamma-ray loud BL~Lac objects are commonly classified as HBLs or
 LBLs on the basis of their ratio of radio to X-ray flux
 i.e. HBLs have a broad-band spectral index $\alpha_{\rm{RX}}
 \leq 0.75$ and LBLs have $\alpha_{\rm{RX}} > 0.75$ \cite{PG95}.
 Consequently, for the LBL sub-class, the synchrotron peak is
 generally observed at IR/optical/UV-frequencies, while the X-ray
 band covers the local minimum of the SED $\nu L_{\nu}$ between
 the two spectral humps. LBLs were thought to represent an
 intermediate object class between FSRQs, which have much higher
 bolometric luminosities, and HBLs, which are the least luminous
 blazers.  In FSRQs $\nu L_{\nu}$ peaks in the IR/optical/UV and
 at MeV-GeV-energies, while in HBLs $\nu L_{\nu}$ peaks at soft
 to medium-energy X-rays and GeV-TeV energies.  The apparent
 strict bimodality in the BL Lac distribution seems, however, now
 to be replaced by a more continuous scenario. New surveys such
 as DXRBS \cite{perl98}, RGB~\cite{LM99} and REX~\cite{cac99},
 which cover previously unexplored regions in the parameter
 space, show a smoother distribution of peak frequencies,
 suggesting that the two classes of objects are not intrinsically
 different, but result from one or a few parameters varying
 within the {\it{same}} source population.

\begin{figure}[hbt]
\epsfig{file=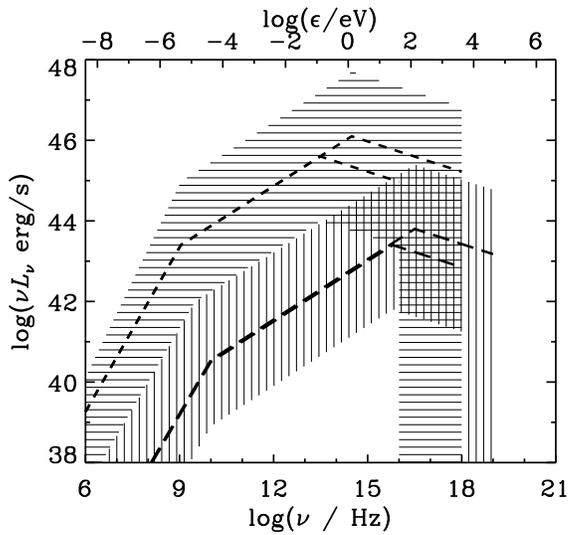, width=10cm}
\caption{Form of the SED assumed for the synchrotron radiation
from LBLs (short dashed curves) and HBLs (long dashed curves).
The horizontal shading encompasses the SEDs of all LBLs, and the
vertical shading encompasses the SEDs of all HBLs considered by
\cite{Gh98}.}
\label{fig:hbl_lbl_sed}
\end{figure}

 In order to identify the critical parameter(s) in the
 framework of the SPB-model we have constructed the ``average''
 synchrotron spectrum for each class, HBLs and LBLs, which serve
 as the target photon distribution for our cascade model. For
 this undertaking we use the ``average'' of an extensive
 collection of blazar SEDs published by Ghisellini et
 al. \cite{Gh98}.  By overlaying all available HBL and LBL SEDs
 we find that the following broken power law gives a reasonable
 representation of the low-frequency hump in the SED of HBLs and
 LBLs
\begin{equation}
n(\epsilon) \propto \left\{ \begin{array}{ll} \epsilon^{-\alpha_1} &
 \mbox{for} \qquad  \epsilon_i\leq \epsilon \leq
 \epsilon_b\nonumber\\
\epsilon^{-\alpha_2} & \mbox{for}
 \qquad \epsilon_b \leq \epsilon \leq \epsilon_c
\end{array} \right.
\end{equation}
 with $\alpha_1=1.5$, $\alpha_2=2.25$ and
 $\epsilon_i=10^{-5}\ldots 10^{-6}$eV.  The break may be
 considered as a result of emission by an electron spectrum that
 is dominated by expansion losses below the break energy, and by
 synchrotron losses above (see e.g. \cite{M93}).  The
 parametrizations for each object class are visualized in
 Fig.~\ref{fig:hbl_lbl_sed}. We find that the break energy
 $\epsilon_b$ of LBLs varies up to about one order of magnitude,
 from $\approx$ 0.1 eV to 1.3 eV, while the maximum synchrotron
 photon energy $\epsilon_c$ can cover a range up to two orders of
 magnitude, from $\sim$ 41 eV to 4 keV. The populated energy
 range of HBLs is more restricted: $\epsilon_b\approx$ 26 eV to
 131 eV and $\epsilon_c\approx$ 4.1 keV to 41 keV. The peak of
 the low-energy SED is $\log{\nu L_{\nu}^{\rm{max}} {\rm
 (erg/s)}} \approx 45.6-46.1$ for LBLs and $\log{\nu
 L_{\nu}^{\rm{max}} {\rm (erg/s)}} \approx 43.4-43.8$ for
 HBLs. We define the ``standard LBL'' by $\epsilon_b=$1.3 eV,
 $\epsilon_c=$4.1 keV and $\log{\nu L_{\nu}^{\rm{max}} {\rm
 (erg/s)}} = 46.1$, and the ``standard HBL'' by $\epsilon_b=$131
 eV, $\epsilon_c=$41 keV and $\log{\nu L_{\nu}^{\rm{max}}} {\rm (erg/s)} =
 43.8$.  In the following parameter study, we use these
 ``definitions'', and vary the magnetic field strength and target
 photon density as the most interesting parameters.


\subsection{Parameter study}

In this section we explore the effects of varying different model
parameters on the blazar SED, especially its high-energy part.

\subsubsection{Magnetic field strength}
\begin{figure}[hbt]
\epsfig{file=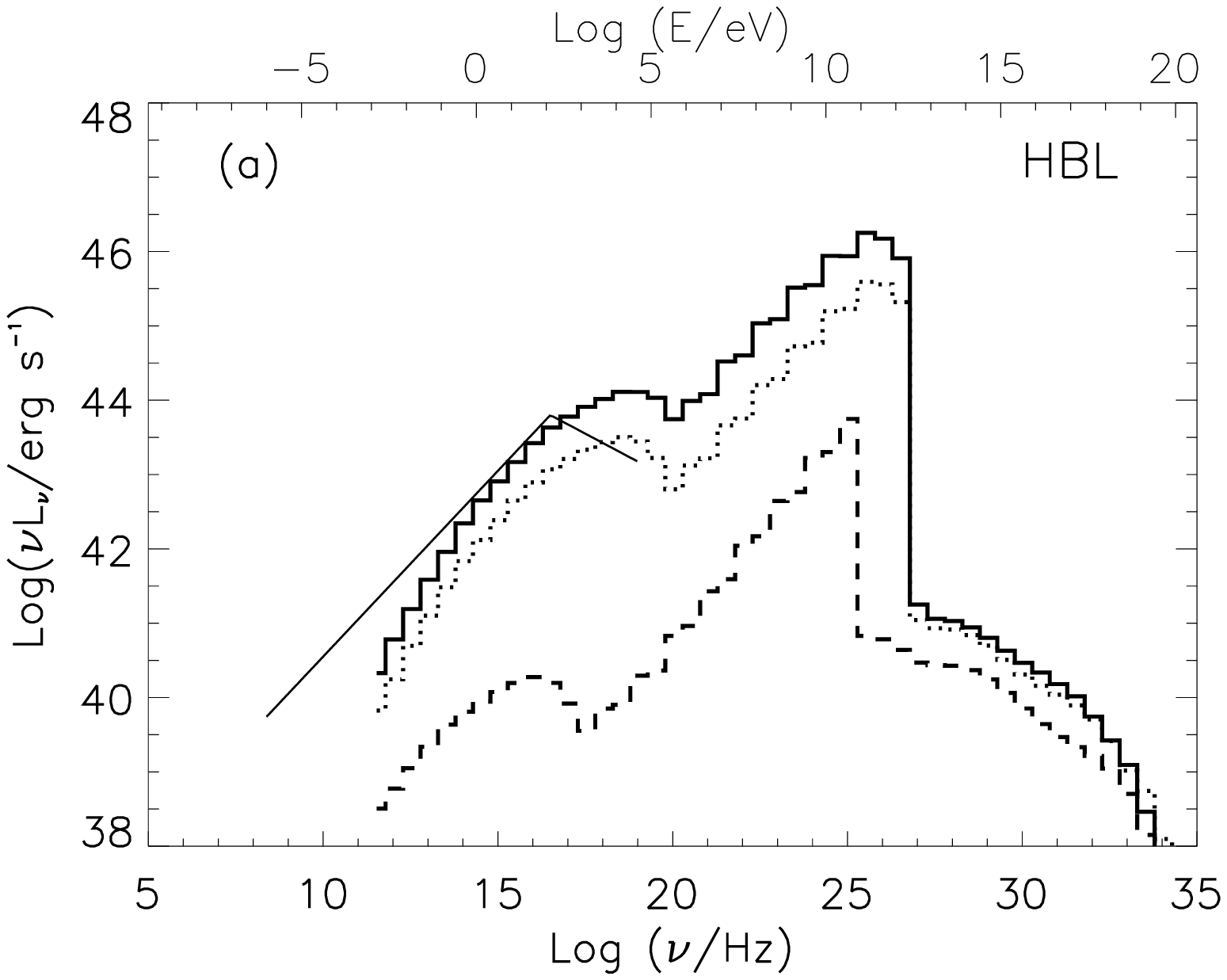, width=7.5cm}
\epsfig{file=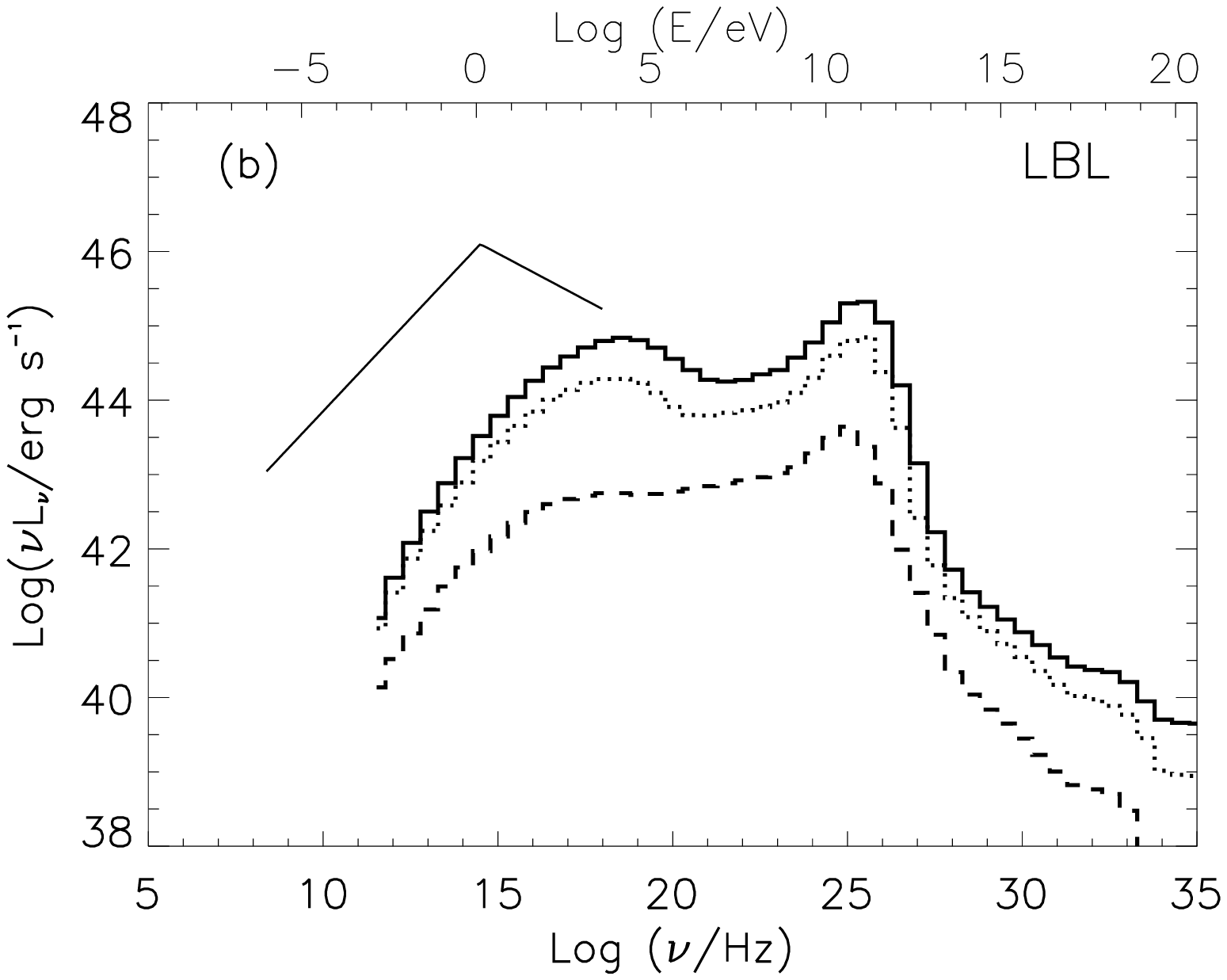, width=7.5cm}
\caption{SED of emerging cascade radiation in the SPB model with
target photon spectra (broken power-law) given by the synchrotron
component of the ``average'' SED in Fig.~1: (a) HBLs with $\nu
L_{\rm{max,syn}} = 10^{43.8}$ erg/s, and (b) LBLs $\nu
L_{\rm{max,syn}} = 10^{46.1}$ erg/s, for $D = 8$, $R' = 5\times
10^{15}$ cm and $B' = 10$ G (dashed histogram), 30 G(dotted
histogram), 50 G (solid histogram).  We assume equipartition
between relativistic proton energy density and magnetic energy
density $u'_B = u'_P$, and this gives $L_{\rm{jet}}/10^{44}$
erg/s $\approx$ 4, 35, 98 for HBLs and LBLs
for the three magnetic fields (see Sect.~5 for the calculation
of $L_{\rm{jet}}$). }
\label{fig:standard_SED}
\end{figure}

 The magnetic field is a key parameter in hadronic models. In
order to accelerate protons to energies above the photopion
production threshold, fields of order 10 G are
necessary. Fig.~\ref{fig:standard_SED} shows the resulting SED
for ``standard'' HBL and LBL target photon spectra in a 10, 30
and 50 G field.  If HBLs possess intrinsically low photon fields,
proton synchrotron radiation always dominates over pion
production. The resulting cascade spectrum
(Fig.~\ref{fig:standard_SED}a) consists mainly of proton
synchrotron emission at gamma-ray energies, and reprocessed
proton synchrotron radiation (i.e. synchrotron radiation from the
pair produced $e^\pm$) at X-ray energies for $B'\geq 10$~G.

 In the denser (LBL-like; see Sect. 3.2.2) photon fields one can
 observe the growing importance of synchrotron losses with
 increasing magnetic field strength. Synchrotron radiation
 generates two distinct humps: one at high energy mainly due to
 the emerging $\mu$ synchrotron radiation, and one at low
 energy dominated by synchrotron radiation of the secondary
 electrons from the $\mu$ synchrotron cascade.  The radiation
 from the $\pi^0$ and $\pi^{\pm}$-cascades adds to the emission
 from the $p$ and $\mu$ synchrotron cascades, and so may fill in
 the gap between the two humps, especially for low magnetic
 fields. For field values below 10~G, the featureless $\pi^0$ and
 $\pi^{\pm}$-cascades may even dominate the emerging cascade
 spectrum. This highlights the need of high field values in
 blazars in the framework of the present model.

 Muon synchrotron radiation, and its subsequent generation(s),
 peaks at higher energies than the corresponding $p$ synchrotron
 radiation~\cite{R99}.  Aharonian \cite{aha2000} has argued that
 a significant flux of TeV-photons requires an optically thin
 emission region at these energies, thus a low target photon
 density and consequently a low efficiency of photopion
 production. In TeV blazars, all of which are HBLs, we therefore
 expect the contribution from the $\mu$ synchrotron cascade to be
 smaller than that from the $p$ synchrotron cascade. (Note, however, that
 the spectral data may be also explainable by models with a moderately
 optically thick photospheric emission in the TeV-regime \cite{M93}
 which would allow a higher photohadronic interaction rate and observable
 emission of $\pi^0$ and $\mu$-induced cascades also in TeV-blazars \cite{R99}.)
 However, it seems to be the opposite in LBLs.  Here, for field values up to
 at least 50 G the $\mu$ synchrotron cascade determines the
 2-peak structure in our chosen ``standard'' LBL, while a
 non-negligible contribution from the $\pi$-cascades fills the
 gap between the humps in the SED, smoothing out the bumpy
 spectral shape.

 We find that with increasing magnetic field, the two-humped
 structure becomes more and more pronounced due to the growing
 $\mu$ and $p$ synchrotron photon production, and the luminosity
 increases partly because we have assumed that the relativistic
 proton energy density increases in equipartition with magnetic
 field energy density.

 The ratio between the low to high energy peak of the cascade
 spectrum is mainly determined by opacity effects, and is in
 general lower in the denser LBL-like target photon fields than
 in the less dense HBL-like environments, as seen in
 Fig.~\ref{fig:standard_SED}.

 The maximum photon energy in the high-energy hump is
 proportional to ${\gamma '}_{\rm{p,max}}^2 B' D$ where ${\gamma
 '}_{\rm{p,max}}$ is the maximum proton Lorentz factor, which in
 turn is determined by balancing the acceleration time scale with
 the energy loss scale, and $D$ is the Doppler factor.
 In HBLs this particle cutoff is caused
 by $p$ synchrotron losses for strong fields, and adiabatic
 losses for fields $B'\leq$ 10 G. In the denser target photon
 fields of LBLs the cutoff occurs at significantly lower
 energies, and is caused by losses through photopion production,
 independently of magnetic field strength.

\subsubsection{Jet-frame target photon field}

 So far, there is no convincing evidence for significantly
 different Doppler factors and emission volumes in HBLs and LBLs.
 As a consequence, the more luminous LBLs
 probably possess larger co-moving frame electron synchrotron
 photon densities than HBLs. In this section we study the effect
 of varying the target photon density on the high-energy hump of
 the SED in the SPB-model.

\begin{figure}[hbt]
\epsfig{file=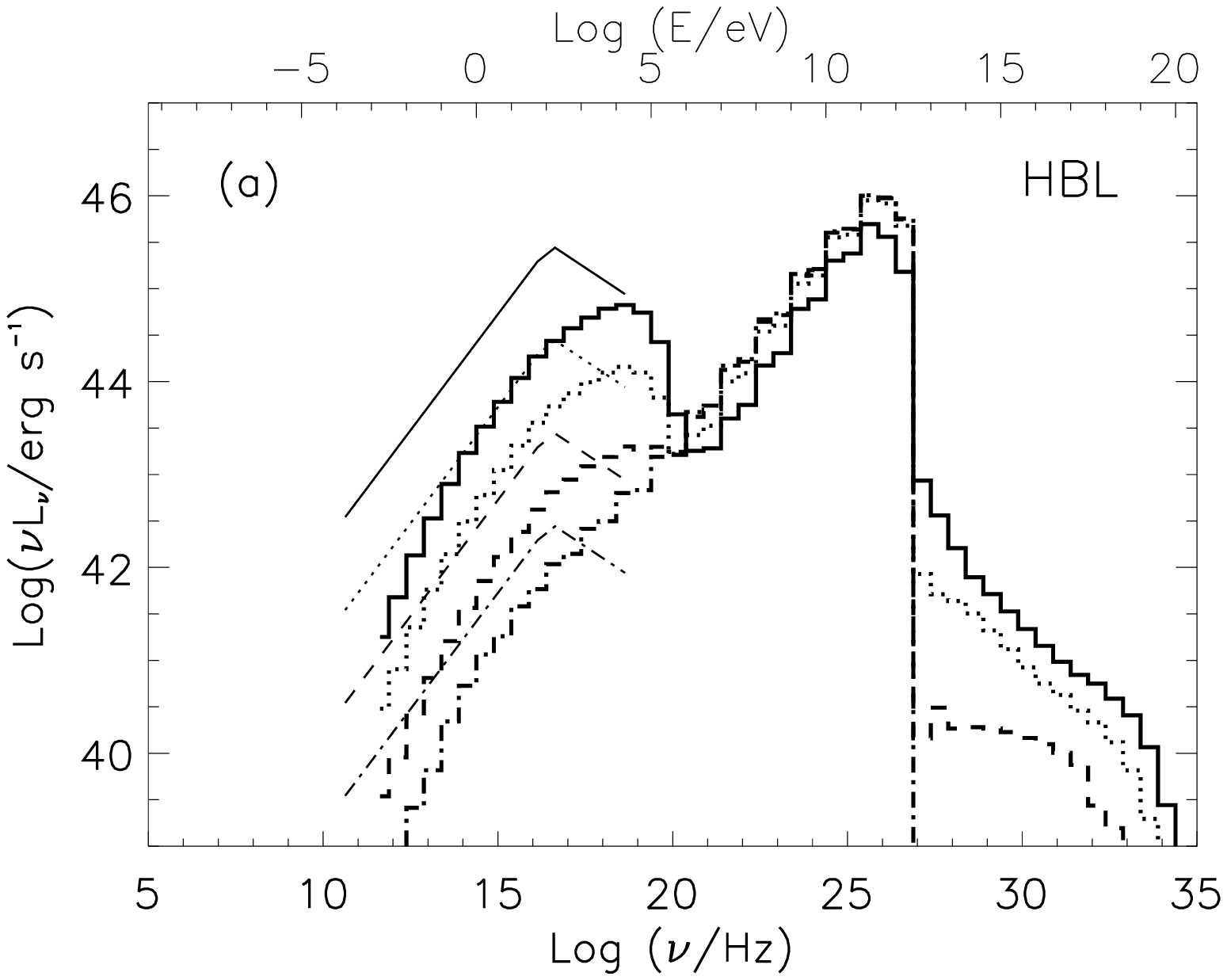, width=7.5cm}
\epsfig{file=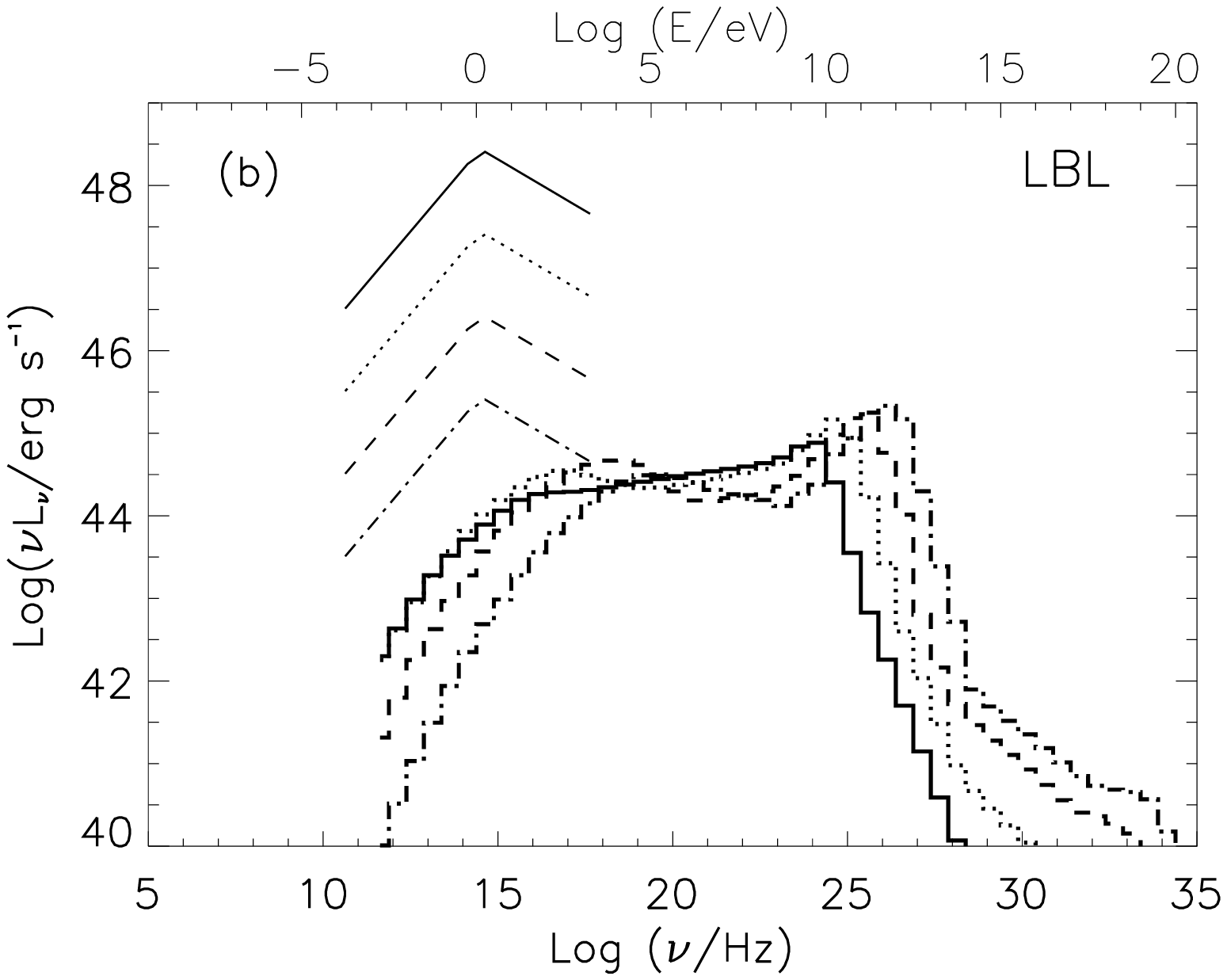, width=7.5cm}
\caption{SED of emerging cascade radiation for different target
photon spectra (broken power-laws shown), $u'_B = u'_P$, $B' = 30
G$, $D = 10$, and $R' = 5\times 10^{15}$ cm.  (a) HBL-like
synchrotron spectra with $u'_B = u'_P$,
$\log(u'_{\rm{phot}}/\rm{eV cm}^{-3}) = 8,9,10,11$ and
$L_{\rm{jet}}/10^{44}$ erg/s $\approx$ 55 for all
$\log(\nu L_{\rm{max,syn}} / \rm{erg s}^{-1}) =
42.4, 43.4, 44.4, 45.4$.
(b) LBL-like synchrotron spectra with
$\log(u'_{\rm{phot}}/\rm{eV cm}^{-3}) = 11,12,13,14$ and
$L_{\rm{jet}}/10^{45}$ erg/s $\approx$ 55, 55, 56, 58
corresponding to $\log(\nu L_{\rm{max,syn}} / \rm{erg s}^{-1}) =
45.4, 46.4, 47.4, 48.4$. }
\label{fig:emerging_cascade}
\end{figure}

 Fig.~\ref{fig:emerging_cascade}a shows the cascade spectrum in a
 HBL-like target photon distribution with varying low-energy
 target photon density indicated in the figure as the broken
 power law curves at the left. The size of the emission region
 $R'$, Doppler factor $D$ and magnetic field $B'$ are held fixed
 at $D=10$, $R' = 5\times 10^{15}$ cm and $B'=30$ G.  For these
 fixed parameters, ${u'}_{\rm{phot}}=10^8$--$10^{11}$
 eV cm$^{-3}$ covers the full range of observed $\nu L_{\nu}$ in
 HBLs as described in Sect.~3.1, and we shall describe the effect
 of varying ${u'}_{\rm{phot}}$ in this range.  As mentioned
 previously, the contribution of $\mu$-synchrotron and
 $\pi$-cascades increases with increasing ${u'}_{\rm{phot}}$
 because of the growing efficiency of photo-meson production.
 This affects not only the shape of the cascade spectrum, but
 also the proton cutoff energy.  Specifically, in
 Fig.~\ref{fig:emerging_cascade}a the proton spectrum cuts off
 due proton synchrotron losses for ${u'}_{\rm{phot}}\leq 10^{11}$
 eV cm$^{-3}$. The corresponding cascade spectra are mainly
 determined by the $p$ synchrotron cascade. For higher
 ${u'}_{\rm{phot}}$ a significant contribution from the
 $\mu$-cascade emission modifies the spectral shape.  The
 apparent increase of the ``dip'' energy (i.e., the energy
 corresponding to the minimum between the low and high energy
 humps of the {\it cascade} spectrum) with increasing
 ${u'}_{\rm{phot}}$ above 10$^{10}$ eV cm$^{-3}$ is caused by an
 increase in the reprocessed $\mu$ synchrotron component.
 Because opacity effects also gain importance for higher photon
 densities, the ratio of the low to high energy peaks of the
 cascade SED increases with increasing ${u'}_{\rm{phot}}$.
 It is
 interesting to note that the soft photon and high energy peak
 luminosities are approximately equal for target densities
 ${u'}_{\rm{phot}}\approx 10^{10}\ldots 10^{11}$ eV cm$^{-3}$, and their ratio
 increases with increasing target photon density.

 Keeping $R'$, $D$ and $B'$ fixed as before, and increasing the
 target photon density even more, one enters the range for
 LBL-like blazars: ${u'}_{\rm{phot}}=10^{11}$--$10^{14}$
 eV cm$^{3}$ would correspond to $\nu L_{\rm max,syn} =
 10^{45}$--$10^{48}$ erg/s. The predicted LBL cascade spectra are
 shown in Fig.~\ref{fig:emerging_cascade}b.  Clearly visible is
 the decrease of the high-energy cutoff with growing
 ${u'}_{\rm{phot}}$. Here, energy gains from acceleration are
 balanced by photoproduction losses in all cases.  Because of the
 dense target photon fields, photo-meson production determines
 not only the cutoff energy but also the spectral shape.  Opacity
 effects are responsible for the steady decline of the high
 energy hump with increasing ${u'}_{\rm{phot}}$ as more and more
 energy is redistributed from high to low energies. The dip
 energy, and corresponding dip luminosity, then become
 increasingly difficult to define.  The
 reprocessed $\mu$ synchrotron radiation, which initially
 dominates the low energy hump of the cascade spectrum, is
 gradually replaced by the featureless $\pi$-cascade emission for
 ${u'}_{\rm{phot}}\geq 10^{11}$eV cm$^{-3}$ which fills the gap
 between the low and high-energy cascade humps.  The $\pi$
 cascade dominates the emission between 1~MeV and 100~MeV for
 ${u'}_{\rm{phot}}=10^{12}$eV cm$^{-3}$, and between 10~keV and
 100~MeV for ${u'}_{\rm{phot}}=10^{13}$eV cm$^{-3}$ in LBL-like
 target spectra.  For denser target photon fields the $\pi$
 cascades completely determine the whole cascade spectrum, and
 proton synchrotron radiation would be unimportant.

 To summarize, in the low target photon densities of HBL-like
 objects $p$ synchrotron radiation and its reprocessed emission
 dominates.  In the very dense target photon environments of
 LBL-like sources photo-meson production becomes so efficient
 that first $\mu$ synchrotron cascade radiation, and finally
 $\pi$-cascade radiation determines the entire cascade SED.

 The radiative efficiency of photopion production is much lower
 than $p$ synchrotron radiation, and this becomes apparent when
 comparing Fig.~\ref{fig:emerging_cascade}a and
 \ref{fig:emerging_cascade}b.  SEDs dominated by $p$ synchrotron
 radiation show in their high-frequency humps power comparable to or higher
 than the seed photon density (mainly generated by synchrotron
 emission from primary electrons).  However, when
 photoproduction is the dominant process, the emerging cascade
 spectra are significantly below the low energy hump caused by
 the directly accelerated electrons.  Thus, if LBL-like objects
 have high density target photon fields $u'_{\rm{phot}}\geq
 10^{11}$ eV cm$^{-3}$, then they would be $\gamma$-ray quiet
 unless one of the other parameters (e.g. $D$, $R'$, $B'$)
 changed in a way to increase the high-frequency hump.  Indeed,
 this could explain why many radio-loud blazars have never been
 observed in $\gamma$-rays.  Intrinsic TeV-emission from LBLs
 could also be suppressed if the particle acceleration efficiency
 $\eta$ is significantly lower than for HBLs.

 The effects of varying the size of the emission region $R'$
 and/or $D$ for a given magnetic field and target photon density
 are obvious: an increase in $R'$ and/or $D$ would result in a
 significant increase in the emerging luminosity, and the maximum
 photon energy would increase with increasing Doppler factor.
 Furthermore, since the photon-photon pair production opacity
 $\tau_{\gamma\gamma} \propto u'_{\rm{phot}} R'$ grows with the
 size of the emission region, it will cause effects similar to
 increasing $u'_{\rm{phot}}$ without, however, any change of the
 $p\gamma$ interaction rate.

\begin{figure}[hbt]
\epsfig{file=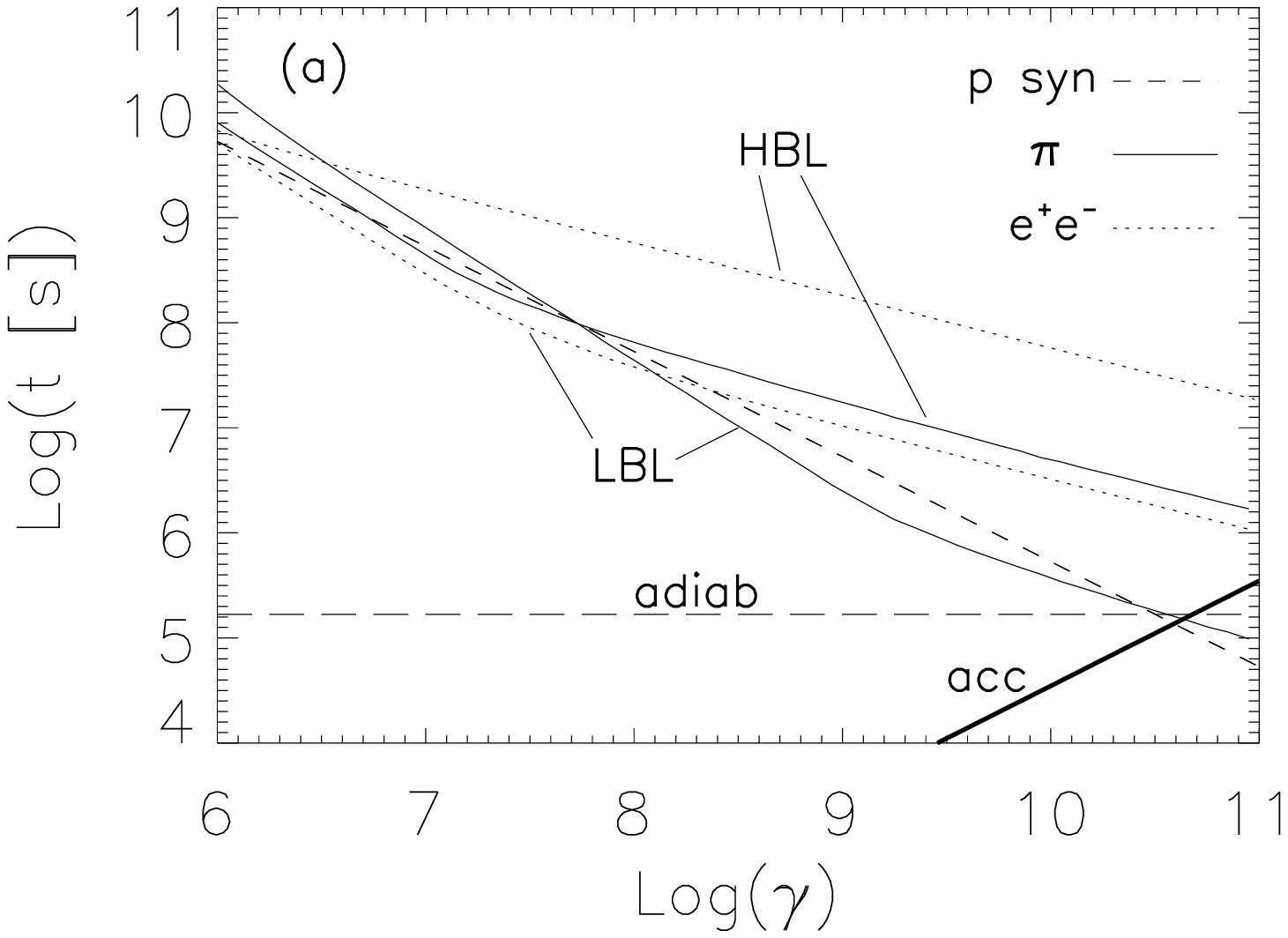, width=7.5cm}
\epsfig{file=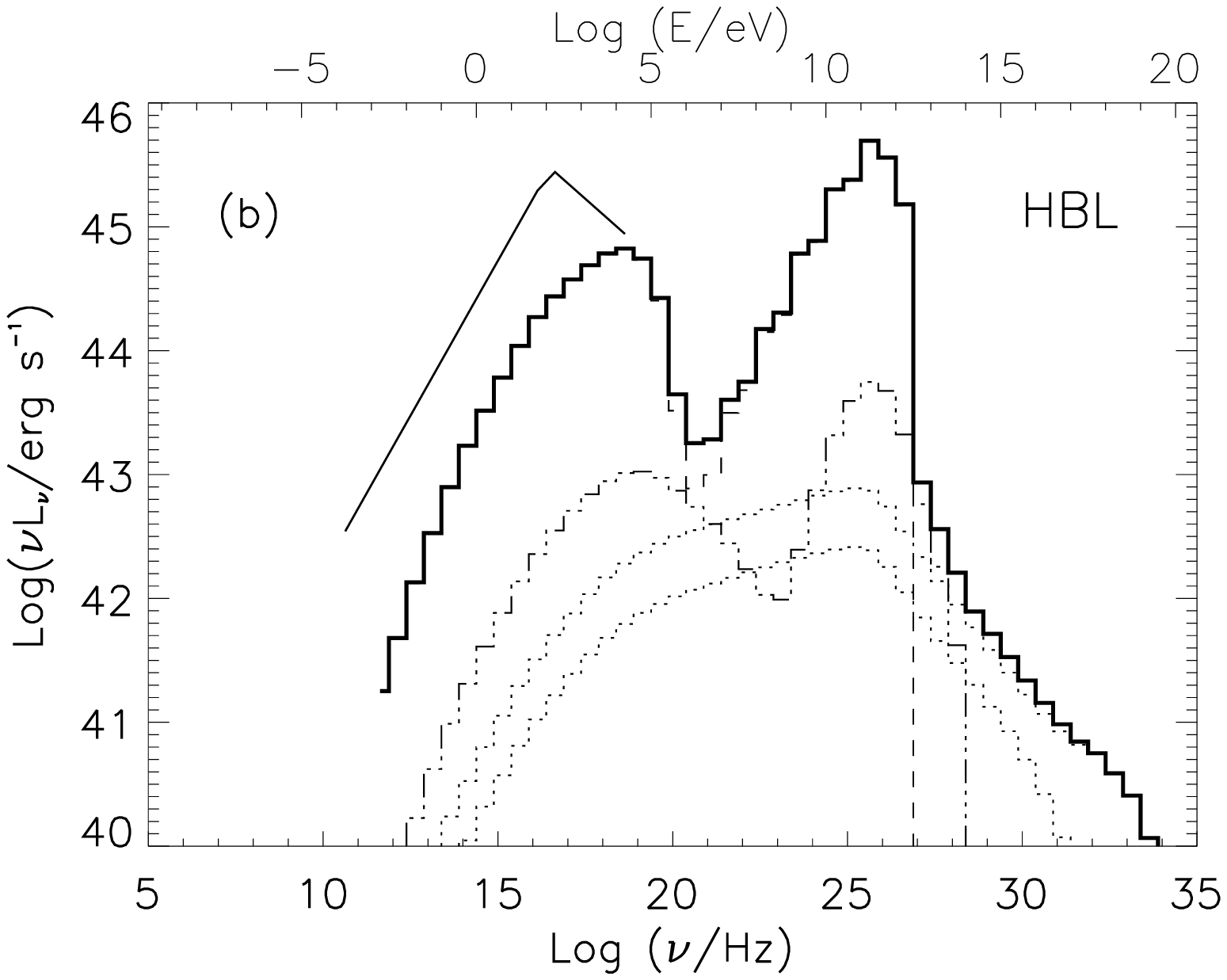, width=7.5cm}
\epsfig{file=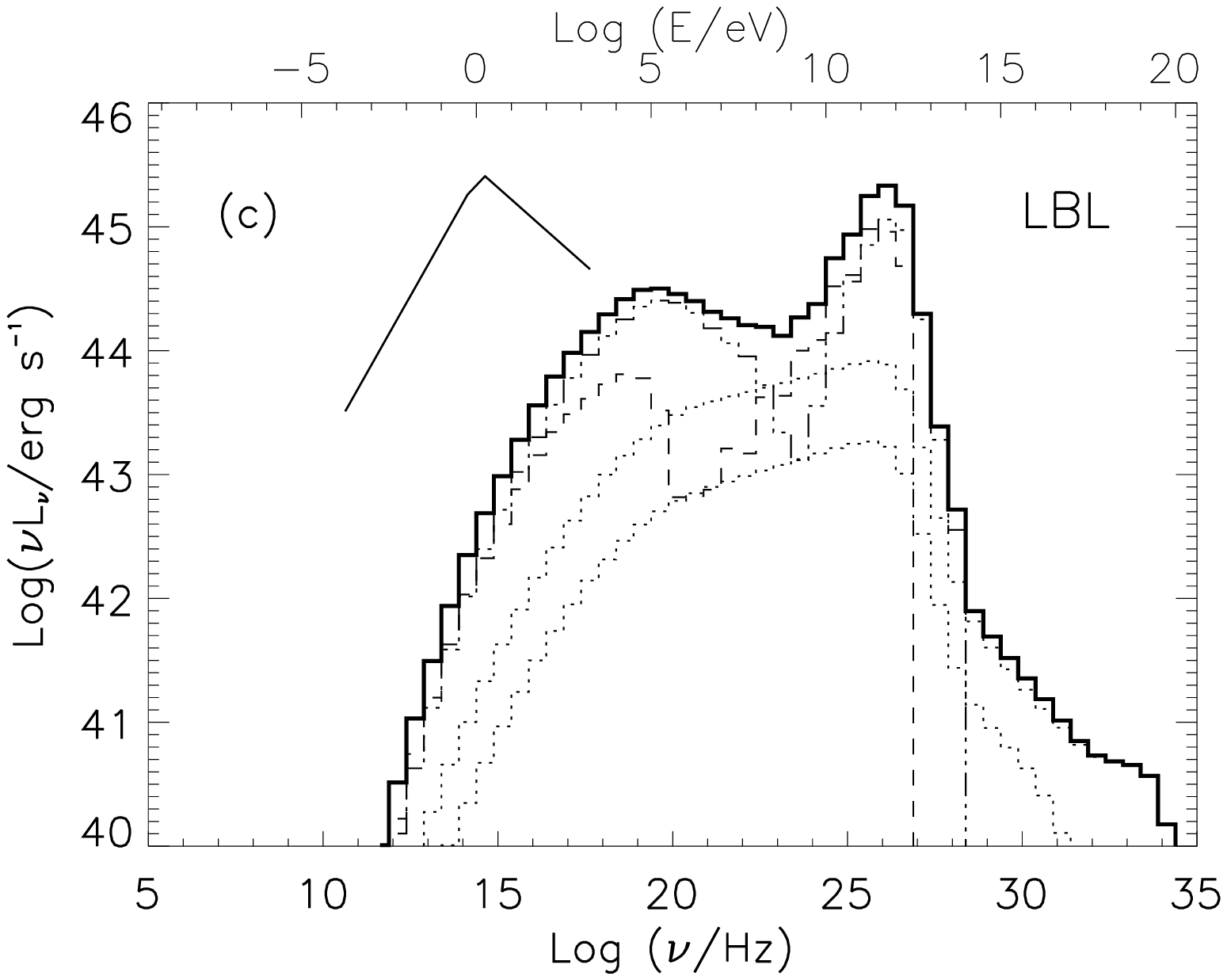, width=7.5cm}
\caption{(a) Mean energy loss time (jet frame) of $p$ in HBL-like and
LBL-like target photon spectra for $\pi$-photoproduction ($\pi$),
Bethe-Heitler pair production ($e^+e^-$) and proton synchrotron
radiation (p syn) for $B' = 30 G$, $\log(u'_{\rm{phot}}/\rm{eV
cm}^{-3})=11$, $D = 10$, $R' = 5\times 10^{15}$ cm, $u'_B =
u'_P$.  The acceleration time scale (acc) is indicated as a thick
straight line. (b) SED of emerging cascade radiation for HBL-like
synchrotron spectra.  (c) SED of emerging cascade radiation for
LBL-like synchrotron spectra.  The target photon spectra are
shown as a broken power-law curves on the left in each figure.
Emerging cascade spectra: $p$ synchrotron cascade (dashed line), $\mu$
synchrotron cascade (dashed-triple dot), $\pi^0$ cascade (upper dotted
line) and $\pi^{\pm}$-cascade (lower dotted line).}
\label{fig:u_11}
\end{figure}

 An interesting case is when proton synchrotron and
 photoproduction losses are approximately equal above the pion
 production threshold.  This occurs, e.g., at target photon densities
 $u'_{\rm{phot}}\approx 10^{11}$ eV cm$^{-3}$ for a 30~G magnetic
 field (see Fig.~\ref{fig:u_11}a). Because the break energy in the low energy
 synchrotron component (used as target for $p\gamma$ interactions
 in our model) is low in LBLs, the pion production loss time
 follows roughly a $\gamma_p^{'-1.25}$ dependence up to $\gamma'_p
 \sim 10^9$, and is at roughly the same level as the proton
 synchrotron loss time $\gamma_p^{'-1}$ from low to high proton
 energies. This leads for the given parameters to approximately
 equal pion production and proton synchrotron losses at high
 energies in LBL. The rather dense target photon field, together
 with a strong magnetic field, are ideal conditions for muon
 synchrotron radiation. Consequently, in LBLs with $B'\approx
 30$~G and $u'_{\rm{phot}}\approx 10^{11}$ eV cm$^{-3}$, the high
 energy hump is mainly due to proton and muon synchrotron
 radiation (see Fig.~\ref{fig:u_11}a). In contrast, HBLs have
 much higher break energies in their low energy synchrotron
 component, with the consequence that proton synchrotron losses
 dominate over pion production losses, leading to a
 high-frequency hump which is predominantly due to proton
 synchrotron radiation (see Fig.~\ref{fig:u_11}b).

\subsection{The LBL/HBL continuity in the SPB model -- a
comparison with the observations}

In this section we propose that HBLs and LBLs are intrinsically
the same objects but with different low energy photon densities
$u'_{\rm phot}$. The photon density may range continuously from low
(HBLs) to high (LBLs) values, resulting in a continuous range of
SEDs in the low energy hump, consistent with the apparent strict
bimodality in the BL Lac distribution being now replaced by a
more continuous scenario as suggested following recent data from
DXRBS \cite{perl98}, RGB~\cite{LM99} and
REX~\cite{cac99}. Because for each parameter set in the SPB model
there exists a $u'_{\rm phot}$ above which $p\gamma$ interactions
dominate, and below which proton synchrotron losses dominate,
this results in a dichotomy when the high energy hump is
included in the SED. Thus, while there is apparently continuity
at low energies, we suggest there is nevertheless a dichotomy at
gamma-ray energies.

\begin{figure}[hbt]
\epsfig{file=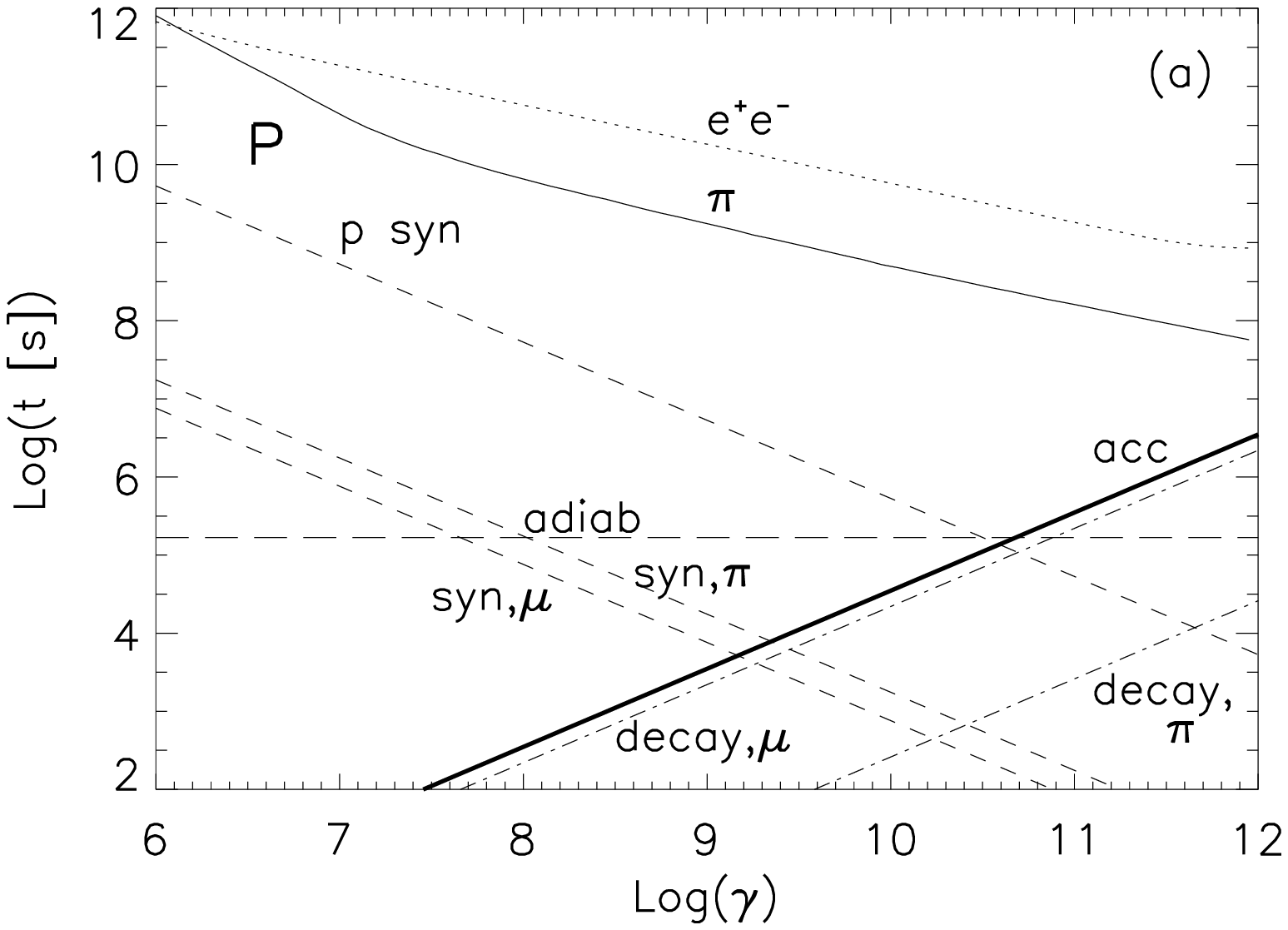, width=7.5cm}
\epsfig{file=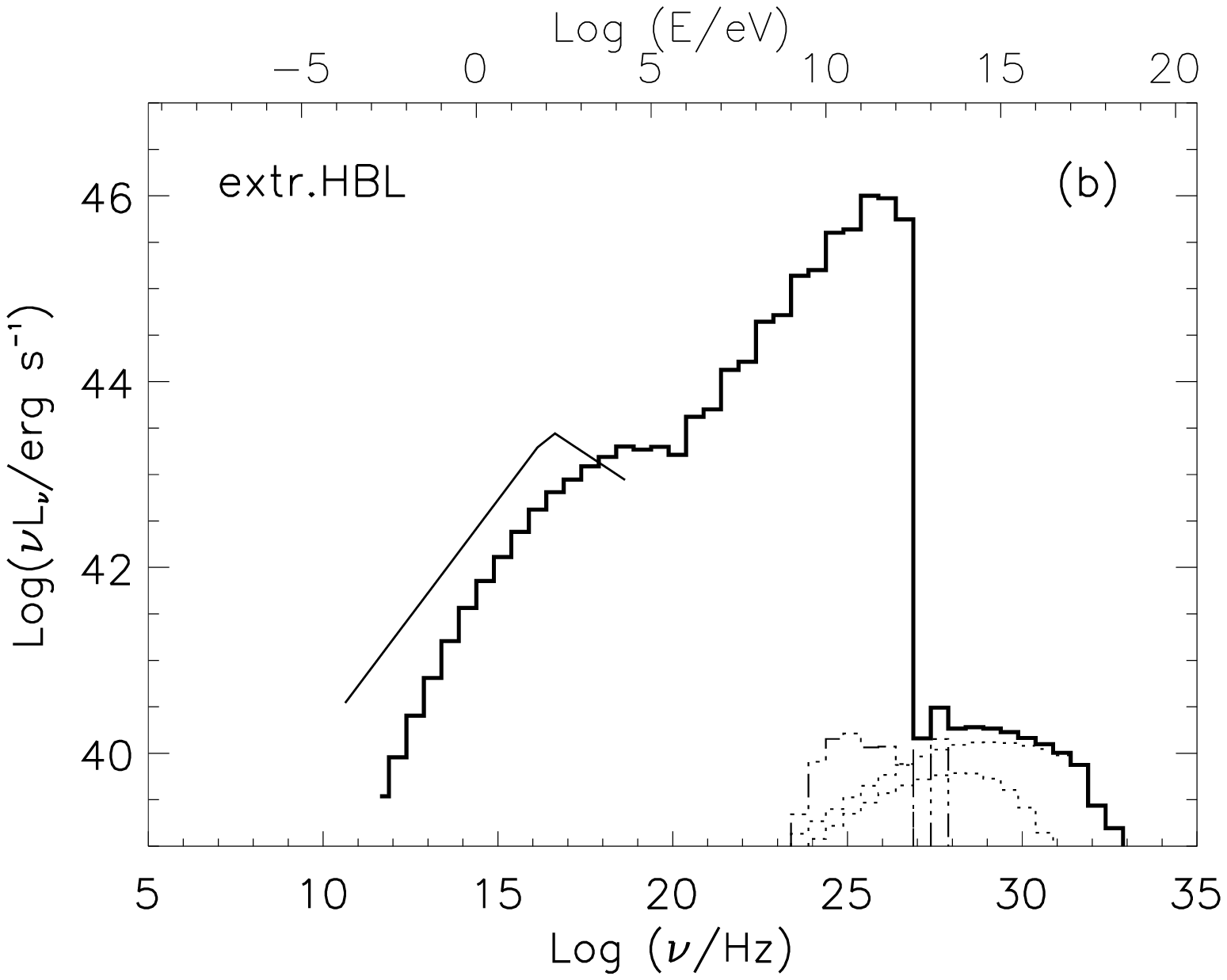, width=7.5cm}
\caption{Example of an extreme HBL. Parameters: $B'=30$ G, $D =
10$, $R'=5\times 10^{15}$cm, $u'_{\rm{phot}} = 10^9$ eV cm$^{-3}$,
$\nu L_{\rm{max,syn}} = 10^{43.4}$erg/s, L$_{\rm{jet}} \approx
5\times 10^{45}$ erg/s, $u'_B = u'_P$, $\gamma'_{\rm{P,max}}= 4 \times
10^{10}$.  (a) Mean energy loss time (jet frame) of $p$ for
$\pi$-photoproduction ($\pi$), Bethe-Heitler pair production
($e^+e^-$) and synchrotron radiation (syn).  Loss times for
$\pi^\pm$ and $\mu^\pm$ for synchrotron radiation (syn $\pi$, syn
$\mu$) are also shown and compared with their mean decay time
scales (decay $\pi$, decay $\mu$). The acceleration time scale
(acc) is indicated as a thick straight line.  (b) Emerging
cascade spectra: $p$ synchrotron cascade (dashed line), $\mu$
synchrotron cascade (dashed-triple dot), $\pi^0$ cascade (upper dotted
line) and $\pi^{\pm}$-cascade (lower dotted line).}
\label{fig:extremeHBL}
\end{figure}

We first discuss an example corresponding to an extreme HBL.
Fig.~\ref{fig:extremeHBL}a shows the relevant time scales for a
relatively low target photon density of ${u'}_{\rm{phot}}\approx
10^{9}$eV cm$^{-3}$, and Fig.~\ref{fig:extremeHBL}b shows the
various cascade components. The clear dominance of the $p$
synchrotron radiation at high energies is obvious, the low energy
component is reprocessed proton synchrotron emission with a peak
luminosity which is significantly lower than that of the
high-energy hump. For much lower target photon fields,
${u'}_{\rm{phot}}\leq 10^{8}$eV cm$^{-3}$, the probability for $p$
synchrotron photons to produce $e^\pm$ pairs is negligible, and
thus reprocessed proton synchrotron radiation does not appear.
Bethe-Heitler pair production and pion photoproduction are also unimportant, as
is the $\mu$ synchrotron cascade.  Because of the low target
photon density in the emitting volume, photons up to several TeV
can escape the emission region.

\begin{figure}[hbt]
\epsfig{file=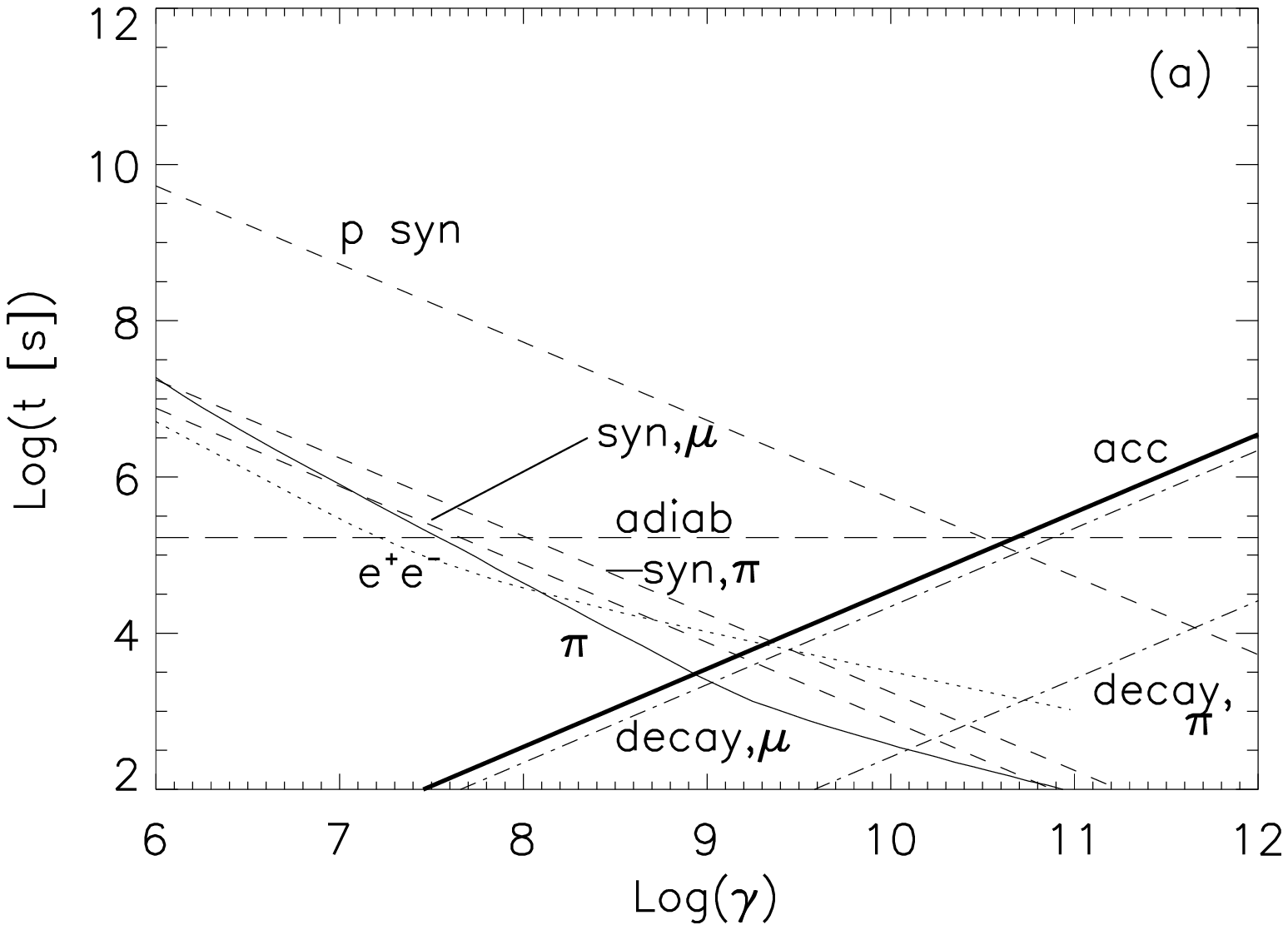, width=7.5cm}
\epsfig{file=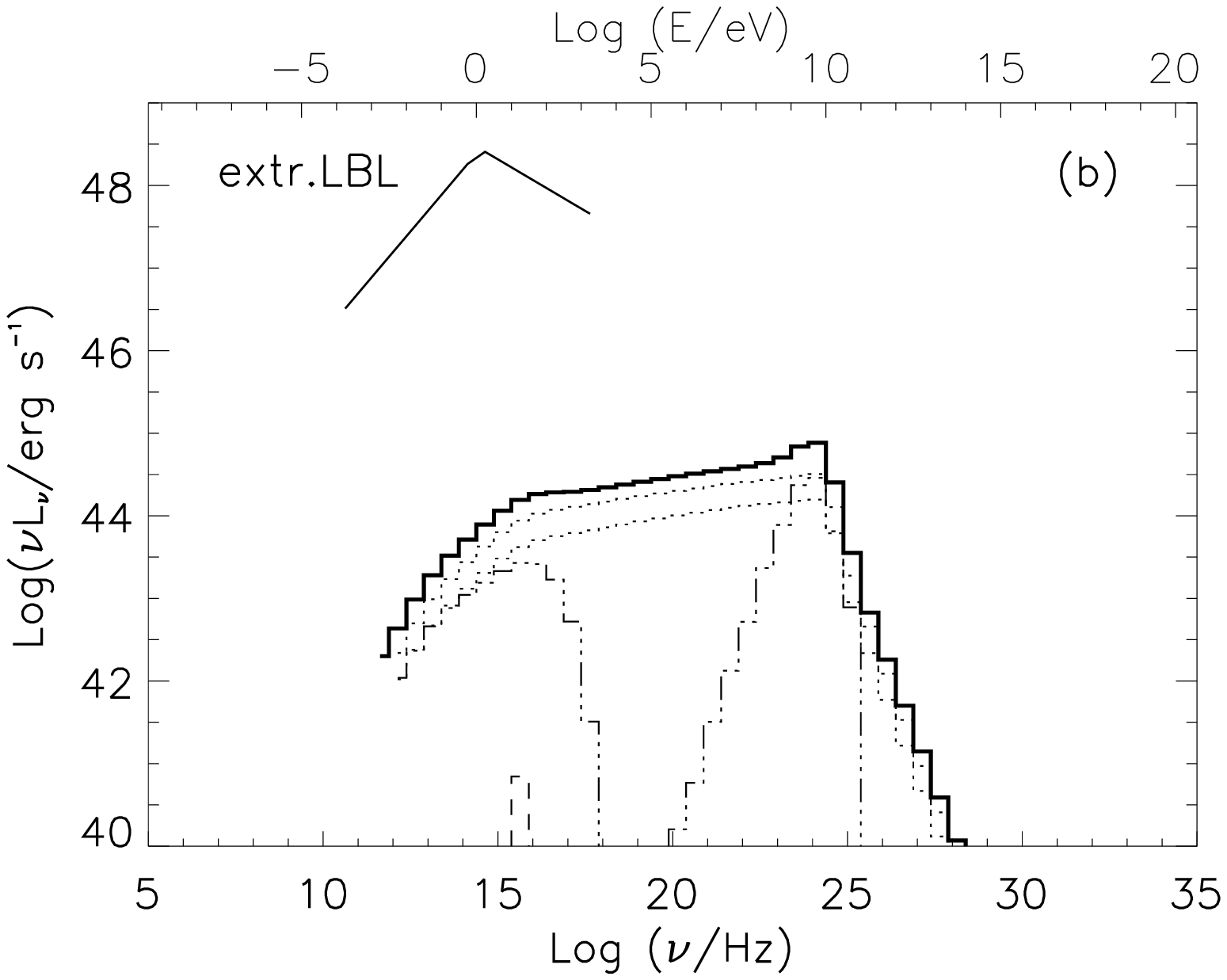, width=7.5cm}
\caption{Example of an extreme LBL: Parameters: $B'=30$ G, $D = 10$,
$R'=5\times 10^{15}$cm, $u'_{\rm{phot}} = 10^{14}$ eV cm$^{-3}$,
$\nu L_{\rm{max,syn}} = 10^{48.4}$erg/s, ${u'}_B$ = ${u'}_P$,
$\gamma'_{\rm{P,max}} = 8\times 10^{8}$,
L$_{\rm{jet}} \approx 6\times 10^{45}$erg/s.
(a) Mean energy loss
time (jet frame) of $p$ for $\pi$-photoproduction ($\pi$), Bethe-Heitler pair
production ($e^+e^-$) and synchrotron radiation (syn).  Loss
times for $\pi^\pm$- and $\mu^\pm$ for synchrotron radiation (syn
$\pi$, syn $\mu$) are also shown and compared with their mean
decay time scales (decay $\pi$, decay $\mu$). The acceleration
time scale (acc) is indicated as a thick straight line.
(b) Emerging cascade spectra: $p$ synchrotron cascade (dashed line),
$\mu$ synchrotron cascade (dashed-triple dot), $\pi^0$ cascade
(upper dotted line) and $\pi^{\pm}$-cascade (lower dotted line).}
\label{fig:extremeLBL}
\end{figure}

Fig.~\ref{fig:extremeLBL} shows an example corresponding to an
extreme LBL, i.e., at the other end of the BL Lac distribution.
Losses due to photo-meson production cut the proton injection
spectrum off at $\gamma'_{\rm{p,max}}\approx 10^9$ (see
Fig.~\ref{fig:extremeLBL}a), with the consequence that only
photons up to several tens of GeV are important in the cascade
spectrum (Fig.~\ref{fig:extremeLBL}b). With $p$ synchrotron
radiation being rather unimportant in this environment, the
cascades initiated by photo-pion production completely dominate
the SED, and produce a rather featureless spectrum, where the dip
between the low- and high energy cascade component has completely
disappeared.

To demonstrate the LBL/HBL dichotomy 
we shall discuss the observed SEDs of a
typical HBL (Mkn~421) and a typical LBL (PKS~0716+714) within the
framework of our SPB model. The data were taken again from
Ref.\cite{Gh98}, and thus represent the average SED for both
objects.  Fig.~\ref{fig:mrk421pks0716}a shows our model fits to
the data (corrected for attenuation during propagation through
the infrared background \cite{BP99}).  Comparing the observed SED
with models shown in Fig.~\ref{fig:emerging_cascade}, it is
immediately apparent that Mkn~421 represents an extreme HBL in
the framework of our SPB model.  Using a relatively low jet-frame
target photon density of ${u'}_{\rm{phot}} \approx 10^{9}$
eV cm$^{-3}$, the high-energy part of the spectrum is completely
determined by $p$ synchrotron radiation. The low-energy hump is
dominated by the synchrotron radiation of the directly
accelerated electrons. 
The shape of the
target spectrum, a broken power law with spectral indices
$\alpha_1=1.5$ and $\alpha_2=2.25$, is the same as used in
Sect.~3.2, but the break energy and normalization has been
adjusted to the data.  In order to explain the emitted energies
of several TeV, the acceleration efficiency must be $\eta\approx
1$.  High values of $\eta$ ($\eta \approx 1$) can realistically
be obtained by diffusive shock acceleration \cite{henri99} or,
even more promisingly, by annihilation of magnetic fields
\cite{haswell92}. A typical Doppler factor of $D\approx 10$ and
emission region radius of $R'=3 \times 10^{15}$ cm has been used
for this fit. 

PKS~0716+714, classified as an LBL by \cite{Gh98}, has been
chosen because of its well-defined low-energy synchrotron
component. Again, the same power law spectral indices are used as
in Sect.~3.2 to represent this component, and the break energy
and normalization are adjusted to the data (see
Fig.~\ref{fig:mrk421pks0716}b). This blazar has been observed to
emit only up to EGRET-energies.  Thus a maximum proton Lorentz factor of
$\gamma'_{\rm{P,max}} \approx 10^{9}$, determined by pion
production losses in our model fit, is enough to explain these
energies which can be reached even with modest
acceleration efficiencies of $\eta \sim 10^{-2}$.

Within the framework of the SPB-model, PKS~0716+714 has a target
photon density of ${u'}_{\rm{phot}} \sim 10^{11}$ eV cm$^{-3}$, and
seems to lie between LBLs and HBLs.  The emission at MeV-energies
is dominated by $p$ synchrotron radiation, while the low-energy
part of the {\it cascade} spectrum is mainly due to reprocessed
$\mu$ synchrotron radiation. The synchrotron radiation from
directly accelerated electrons is, however, responsible for the
observed low-energy hump. High energy photons up to multi-TeV are
expected from PKS~0716+714 due to $\mu$-synchrotron
radiation. This would be detectable by high-sensitivity
Cherenkov-telescopes such as MAGIC and VERITAS if
photon-absorption in the cosmic background radiation field would
be negligible for this source, but the unknown redshift (we have
assumed here $z=0.3$ \cite{Gh98}) makes this prediction rather
uncertain. However, the MeV-GeV emission predicted from
PKS~0716+714 should be definitively detectable by GLAST \cite{GLAST}.


\subsection{From quiet to flare state in the SPB model}

Variability in the present model could be caused by an increase
in the accretion rate causing a shock to propagate along the jet.
Pre-existing density enhancements or ``blobs'' could thereby be
re-energized, and may undergo an increase in their bulk Lorentz
factor.

To demonstrate how this could work in HBLs, we have chosen the
$\gamma$-ray loud BL Lac object Mkn~501 where the different stages
of activity seem to be sampled best
\cite{pian98}. Fig.~\ref{fig:mrk501flare} shows our modeling of
the three main activity states (historical, 7 April 1997 and 16
April 1997) leading to the giant 1997-flare. Mkn~501 is an HBL,
and so the dominant $\gamma$-ray production mechanism in our SPB
model should be proton synchrotron emission.  This would be the
case even for the quiet state, although we note that the
high-energy component in this state is not very well constrained,
and other parameter choices can easily be found to fit the data
equally well. Future data from more sensitive instruments,
e.g. GLAST \cite{GLAST}, may provide better constraints.  For
each of the three states, we show the target photon spectrum
(solid curves) we use for pion photoproduction, and cascading.

As the shock moves through the highly magnetized plasma,
electrons start to increase their synchrotron photon production,
possibly due to an increase in the number of relativistic
electrons.  This leads to a higher {\it{intrinsic}}
(i.e. jet-frame) photon density, and thus to an increase of the
observed synchrotron hump, from $\nu L_{\nu,{\rm{max}}} =
10^{44.2} \mbox{ to } 10^{44.7} \mbox{ to } 10^{45.4}$ erg/s for
the Mkn~501 1997-outburst. Simultaneously, the number of
relativistic protons also increases.
The appearance of a "fresh" relativistic shock in an otherwise weakly turbulent
plasma implies qualitatively an increase of the acceleration efficiency $\eta$.
In our model we increased $\eta$ from $\eta=0.05$ to $\eta=1$
during flaring, and this naturally leads to a higher and
correlated cutoff energy of both electron and proton particle
spectra, and consequently to a {\it{correlated}} shift of
synchrotron peak energy and $\gamma$-ray peak energy to higher
energies. We also note that the observed break in the electron
synchrotron spectrum becomes less pronounced with flaring
activity, as is expected when the energy loss time scale becomes
comparable to the dynamical (i.e.\ light-crossing) time scale for
a particle injection spectrum steeper than $\alpha_p \geq 1$
\cite{aha2000}. For simplicity, however, we have used a broken
power law spectrum with the same break (i.e.\ $\alpha_1=1.5$,
$\alpha_2=2$) for all three activity stages to represent the
target photon spectrum for our SPB-model.  The target photon
spectrum break and cutoff energies used were $\epsilon_b=2$~eV
and $\epsilon_c=4$~keV for the quiescent state,
$\epsilon_b=1$~keV and $\epsilon_c=160$~keV for the intermediate
state, and $\epsilon_b=40$~keV and $\epsilon_c=200$~keV for the
flaring state.

Due to the higher acceleration efficiency, the cutoff energy in
the injected proton spectrum, determined by synchrotron
losses, increases by about a factor of 5 in our example. Although
the jet frame target photon density increases nearly a factor of
five from the quiescent stage to the flare stage, photo-meson
production, and thus $\mu$ synchrotron emission as proposed in
\cite{R99} to explain the Mkn~501 flare, remains still of minor
importance.

The combined effect of increasing proton cutoff energy, and
increasing Doppler factor, causes an increase of the ``dip''
energy in the blazar SED. This behaviour is also expected for
leptonic SSC models (see \cite{pian98}), as a result of the
increasing synchrotron and Compton peak energy and the flattening
of the underlying electron spectrum.

The magnetic field and the jet frame emission volume remain
approximately constant during an outburst. Thus the apparent
luminosity rise during an outburst is mainly caused by an
increase of beaming. Because protons are co-accelerated with the
electrons, our model predicts correlated variability of the
synchrotron and high-energy SED component, with a possible lag of
the high-energy hump caused by the longer acceleration and energy
loss time scale of the protons in comparison to the electrons.


\section{Predicted neutrino emission}

\subsection{Neutrino spectra}

In contrast to leptonic models, in hadronic models $\gamma$-ray
production by pion photoproduction would result in simultaneous
neutrino production. The main neutrino production channel is
through the decay of charged pions, e.g.  $\pi^\pm\to\mu^\pm
+\nu_{\mu}/\bar{\nu_{\mu}}$ followed by $\mu^\pm\to e^\pm +
\nu_e/\bar{\nu_e} + \nu_{\mu}/\bar{\nu}_{\mu}$. The neutrinos
escape without further interaction.  Fig.~\ref{fig:neutrino}
shows the predicted average neutrino emission from Mkn~421 and
PKS~0716+714. The photon-hadron interactions for both, LBLs and
HBLs, take place predominantly in the resonance region. Here,
$\pi^-$ and thus $\bar{\nu_e}$ production is suppressed.  We
give the predicted neutrino emission from the objects themselves,
and do not consider here any additional contribution from
escaping cosmic rays interacting while propagating through the
cosmic microwave background radiation.

Provided Mkn~421 and PKS~0716+714 are typical for their
respective object class, and that the Doppler factor in Mkn~421
is comparable or higher than in PKS~0716+714, we find a clear dominance of neutrino
emission from LBLs in comparison to HBLs. The reason for this is
the higher meson production rate in the LBL source population due
to their higher target photon fields in comparison to the HBL
population where proton synchrotron losses dominate. Thus more
power, in our modeling for Mkn~421 by a factor of $10^{3}$--$10^{4}$, is
channeled into the photon component in HBLs, while in LBLs the
power output of photons and neutrinos is approximately
equal. This is in contrast to previous hadronic jet models which
predict equal photon and neutrino energy fluxes for all blazar
types (e.g.\ \cite{M93}).

The neutrino production spectrum depends on the ambient proton
spectrum and the spectrum and density of target photons.  The
proton injection spectrum is modified by interactions and energy
losses.  For Mkn~421 the photopion production rate approximately
follows a broken power law, $t_{\pi}^{'-1} \propto \gamma_p^{'1.25}$
for proton energies below $\sim 10^{7}$~GeV, and $t_{\pi}^{'-1}
\propto \gamma_p^{'0.5}$ above $E_p' \sim 10^{7}$~GeV due to the
break in the target photon spectrum (see Fig.~1). This leads to a
break in the neutrino spectrum at $\sim 10^{7}$~GeV (observer
frame), from power spectral index $\alpha_\nu \approx 1.25$ to
$\alpha_\nu \approx 0.5$ where $(E^2 dN/dE) \propto
E^{\alpha_\nu}$. A further low energy break in the neutrino
spectrum at $\sim 10^4$ GeV is caused by the change of photopion
production rate at threshold. The high-energy cutoff at $\sim
10^9$ GeV in the observer's frame, caused by
$\mu^\pm$-synchrotron losses ($\pi^\pm$-synchrotron emission is
unimportant), is below the strict upper limit of
$E_\nu \leq 3\times 10^{10}$ GeV predicted by \cite{RM98}.
Another important source of high energy neutrinos
is the production and decay of charged kaons when the
proton-photon interaction takes place predominantly in the
secondary resonance region of the cross section
\cite{SOPHIA}. This might be the case for HBLs because their
target photon field can extend up to X-ray energies.  Positively
charged kaons decay with $\sim 64\%$ probability into muons and
direct high energy muon-neutrinos. These muon-neutrinos will not
have energies reduced as a result of synchrotron radiation by
their parent particles. Unlike the neutrinos originating from
$\pi^\pm$ and $\mu^\pm$-decay, they will dominate the
neutrino flavors at the high energy end of the emerging neutrino
spectrum ($E_\nu > 10^{9}$~GeV), and also cause the total
neutrino spectrum to extend to $\sim 10^{10}$~GeV.

We expect the neutrino emission from PKS~0716+714 to be cut off
at $\sim 10^9$ GeV (observer frame) for all neutrino flavors (see
Fig.~\ref{fig:neutrino}) due to a roughly one order of magnitude
lower proton cutoff. Also $\mu^\pm$ synchrotron losses may play a
role here, and are expected to cut off at the same neutrino energy
of $\sim 10^9$ GeV.  The neutrino spectrum follows a power law
with index $\alpha_\nu \approx 1.25$ below the cutoff, and is
caused by photohadronic interactions with the target photon field
above $\epsilon_b'$. Because of the $\pi$-production threshold
and the relatively low proton cutoff in LBLs, meson production in
the photon field below $\epsilon_b'$ cannot occur.

In addition to the neutrinos from the meson decay chain, there
 will be a small contribution of $\bar{\nu}_e$ from neutron
 decay (not shown in the figures). For $R'$ of 10$^{15}$ to 10$^{17}$ cm only neutrons of
 energy 30 to 3000 GeV will decay inside the production
 region.
High-energy neutrons leaving the source into the observer's
sight line will decay and produce highly beamed neutrinos,
 which also makes them appearing to come from the AGN.
In addition, there will be also a contribution from neutrons
decaying close enough to the system to appear a point source to
the observer.
 The $\bar\nu_e$ flux from neutron decay at about 1 TeV
 will be approximately the same as the $\nu_e$ flux at that
 energy.

Neutrons produced during pion photoproduction interactions of protons
may escape from the emission region if the optical depth for
neutron-photon inetractions, $\tau_{n\gamma}$, is less than one.  In this
optically thin case, the escaping neutrons may decay outside the AGN to
become cosmic rays.  Since the output of cosmic rays would then be
related to the
neutrino output, the observed cosmic ray intensity can be used to set an
upper bound on the extragalactic neutrino background (ENB) due to neutrino emission by optically thin
photoproduction sources \cite{WB99,MPR01}.  If, on the other hand, the source is
optically thick to neutron-photon interactions the upper bound on the
ENB intensity \cite{MPR01} is obtained instead from the observed diffuse gamma-ray
background. We can determine which bound, optically thick or optically thin, should
apply to LBLs and HBLs by considering the relevant time scales.
The neutron-photon optical depth is $\tau_{n\gamma}=(R'/c) t'_{n\gamma}$.
Note that since $t'_{\rm{ad}} \approx R'/c$ (Appendix A) and that to a reasonable
approximation $t'_{n\gamma} \approx t'_{\pi}$, we have
$\tau_{n\gamma} \approx t'_{\rm{ad}}/t'_{\pi}$.  From Figs. 4a, 5a and 6a we see that
for HBLs $\tau_{n\gamma} \ll 1$, and that for LBLs $\tau_{n\gamma} > 1$.  Thus the
optically thin bound should apply for HBLs and the optically thick
bound for LBLs.

\subsection{Diffuse neutrino fluxes}

From Fig.~\ref{fig:neutrino} we expect that the contribution from
LBLs to the diffuse extragalactic neutrino background (ENB) would
dominate unless HBLs turn out to be significantly more numerous
than LBLs.  There are several methods to estimate the diffuse
neutrino background.  One method which has been suggested
in \cite{MP2001b} is to normalize to the observed extragalactic
diffuse $\gamma$-ray background \cite{sree98}. If LBLs dominate
the blazar contribution of $\sim 25\%$ \cite{CM98} to the
extragalactic $\gamma$-ray background at GeV energies, then we
would expect a diffuse integrated neutrino intensity of $\approx
10^{-7}$ GeV cm$^{-2}$ s$^{-1}$ sr$^{-1}$.  However, HBLs may
dominate the TeV-photon background, and by extrapolating the
observed photon background to TeV-energies and normalizing the
neutrino output from HBLs to the TeV-background, we would find a
diffuse flux from HBLs of about 3--4 orders of magnitude
lower. Taking this into account, the predicted ENB intensity from
HBLs lies several orders of
magnitude below the neutrino upper bounds \cite{WB99,MPR01}.

If the LBL and HBL luminosity function is known, the diffuse
neutrino background is readily derived by convolving this
luminosity function with the neutrino output of each source and
integrating over redshift:
$$
I_{\rm{ENB}}(E) = \frac{1}{4\pi}\int dz \frac{dV_c}{dz} \frac{1+z}{4\pi d_L^2}
\int dL^{\rm{peak}} \rho(L^{\rm{peak}},z) I_\nu((E(1+z),L^{\rm{peak}})
$$
where $V_c$ is the co-moving cosmological volume, $z$ is the
redshift, $d_L$ is the luminosity distance, and $\rho(L,z)$ is
the source differential luminosity function at peak luminosity,
$L^{\rm{peak}}=\nu L_{\nu}^{\rm{peak}}$ of the low energy hump of
the SED.  We have assumed no luminosity or density evolution for
BL Lac Objects, in good agreement with the observations
\cite{pado01,giom01,Cacc01}. Note that the apparent negative
evolution of HBLs found in early studies might be a selection
effect \cite{giom01}. For the LBL differential BL Lac luminosity
function we used the 5~GHz luminosity function of Urry et al.\
\cite{UPS91} which we parametrize as
$$
\rho (L_{\rm{5GHz}}) =
10^{-33}\left\{\left[\left(\frac{L_{\rm{5GHz}}}{L^*}\right)^{-1.6}\right]^{-1}
+\left[\left(\frac{L_{\rm{5GHz}}}{L^*}\right)^{-3.3}\right]^{-1}\right\}^{-1}
\rm{Gpc}^{-3} (\rm{erg \, s^{-1} \,Hz^{-1}})^{-1},
$$
where $L^*=2\times 10^{33}\rm{erg \, s^{-1} \,Hz^{-1}}$.
We use $q_0 = 0$ and $H_0 = 50$km s$^{-1}$ Mpc$^{-1}$ as in
\cite{UPS91}. The total radio BL Lac space density is estimated to be $\sim
1000$~Gpc$^{-3}$ with the HBL contribution being about 10\% of
the LBL contribution, and so we used
$N_{\rm{HBL}}/N_{\rm{LBL}}=0.1$ in our calculation.
Fig.~\ref{fig:hbl_lbl_sed} is used to convert the 5 GHz radio- to
the synchrotron peak luminosity.  For an LBL with $\nu
L_{\nu}^{\rm{peak}}$ at 1.3 eV ($3.15\times 10^{14}$ Hz), such as
in PKS 0716+714, and assuming that $\nu L_\nu \propto \nu^{0.5}$
for $\nu=5$~GHz, then $\nu L_{\nu}^{\rm{peak}}/\nu L_{\rm{5 \,
GHz}} \approx 250$.  The range of peak luminosities covered by
the luminosity function is simulated by using different jet frame
target photon densities, assuming that the directly accelerated
$e^-$ are responsible for the synchrotron hump and also represent
the target photon field for $p\gamma$-interactions and cascading.
We keep all other parameters at the values derived for
PKS~0716+714.  A similar procedure is used for the HBLs where the
parameters for calculating the $\nu$-spectra are derived from the
fit to Mkn~421, and we used a range of target photon densities to
simulate the luminosity range covered by the predicted luminosity
function.  For LBLs we used $u'_{\rm{phot}}\approx 10^5 \ldots
10^{12}$eV cm$^{-3}$, and for HBLs we have $u'_{\rm{phot}}\approx
10^7 \ldots 10^{12}$eV cm$^{-3}$.  The cut-off energy of the
injected proton spectrum is calculated self-consistently for each
$u'_{\rm{phot}}$ value, and is mainly due to pion production
losses for LBLs, and due to proton synchrotron losses for
HBLs. The resulting diffuse neutrino fluxes are shown in
Figs.~\ref{fig:diffuse}a,b as the solid curves bounding the upper
shaded region, where we have integrated over redshift up to $z=3$.
Also shown in these figures is the cosmic ray induced neutrino
bound derived by \cite{WB99,MPR01} for optically
thick (Fig. 10a) and thin (Fig. 10b) sources
and no source evolution.

\begin{figure}[hbt]
\epsfig{file=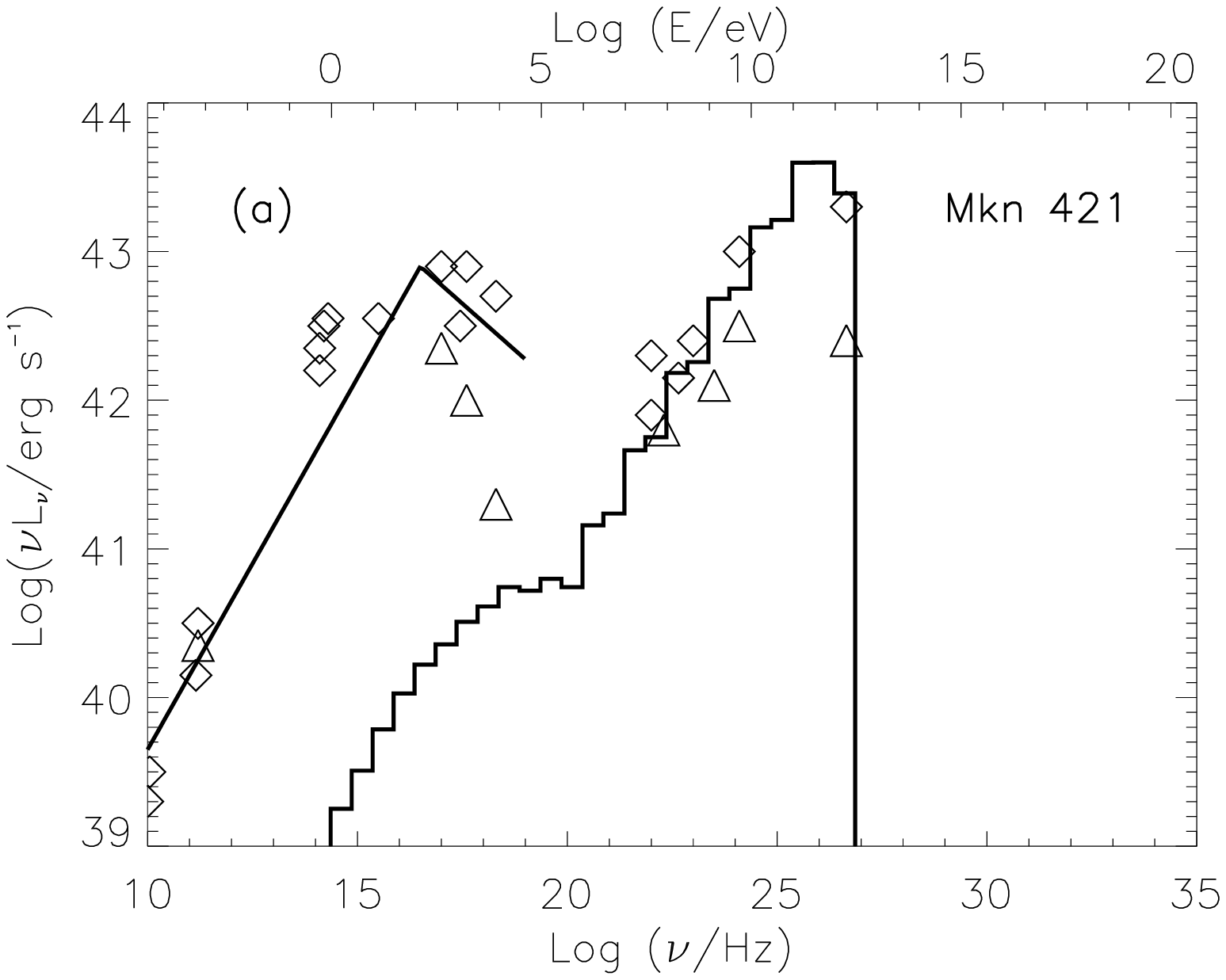, width=7.5cm}
\epsfig{file=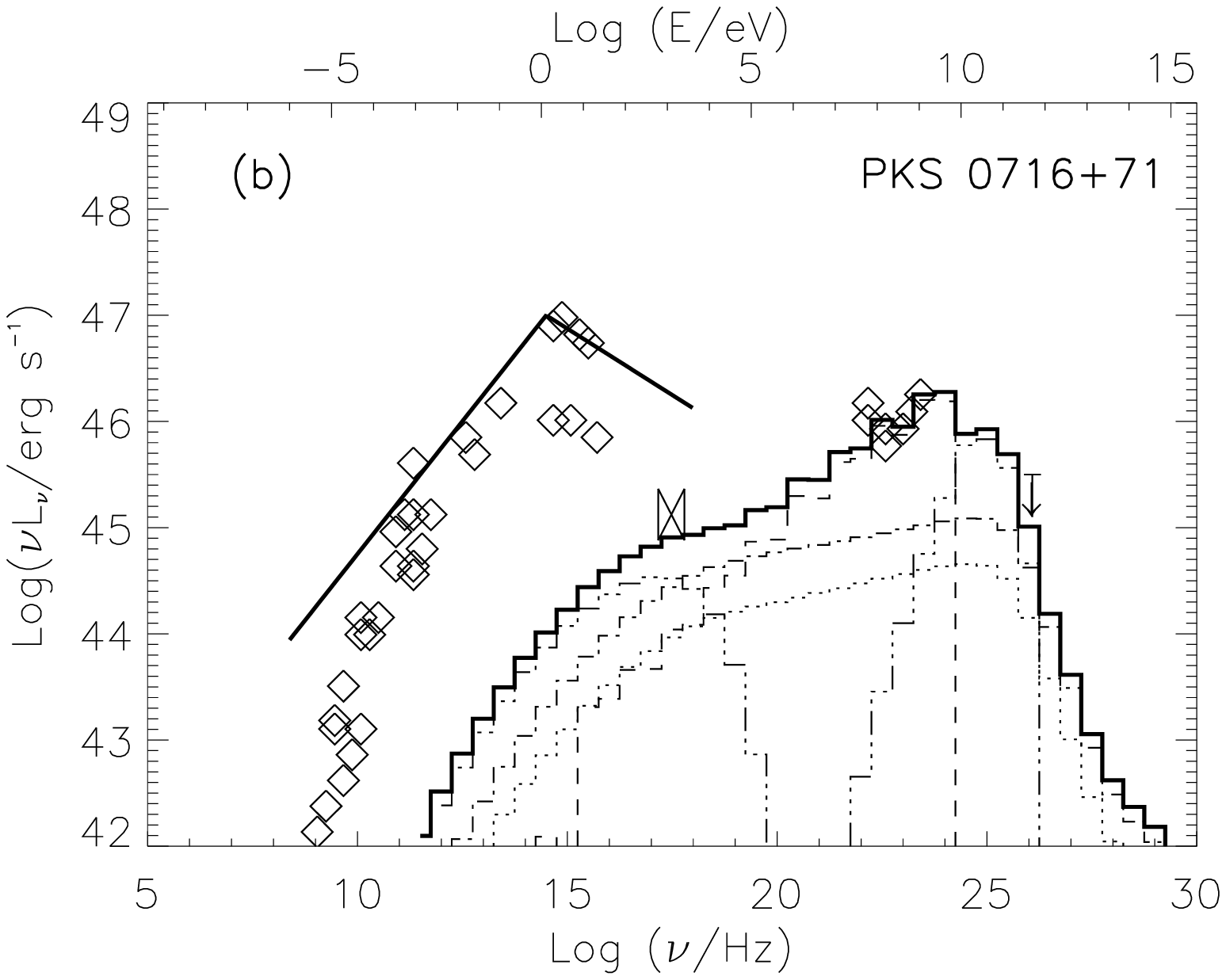, width=7.5cm}
\caption{Modeling the SED of (a) Mkn~421 and (b) PKS~0716+714.
The Mkn~421 data are corrected for pair production on the cosmic
background radiation field \cite{BP99}, and represent the emitted
spectrum at the source.  The target photon distribution is
indicated as a broken power law on the left side in each
figure.
Dotted line represents the $\pi^0$-component, dashed-dotted line the
$\pi^{\pm}$-component, dashed-triple dot the $\mu$ synchrotron, dashed
line the $p$ synchrotron component, and the solid line is the sum of all
cascade components.
Model parameters are: (a) B'=30 G, D = 10, $R'= 3\times
10^{15}$ cm, $u'_{\rm{phot}} = 10^{9}$ eV cm$^{-3}$, $u'_P\approx
1$ erg/cm$^3$, $\gamma'_{\rm{P,max}} = 4 \times 10^{10}$,
$L_{\rm{jet}} \approx 9\times 10^{44}$ erg/s,
$\eta = 1$. (b) $B'=30$ G, $D\approx 7$, $R'\approx
10^{17}$ cm, $u'_{\rm{phot}} \approx 10^{11}$ eV cm$^{-3}$, $u'_P
\approx 6$ erg/cm$^3$, $\gamma'_{\rm{P,max}} = 3 \times
10^{9}$, $L_{\rm{jet}} \approx 3\times 10^{47}$
erg/s, $\eta = 0.01$. The $3\sigma$ upper limit is from
1994-Whipple observations \cite{kerr95}. Note that the absorption
effects in the cosmic background radiation field are not taken
into account here because of the uncertain redshift of
PKS~0716+714.}
\label{fig:mrk421pks0716}
\end{figure}

\begin{figure}[hbt]
\epsfig{file=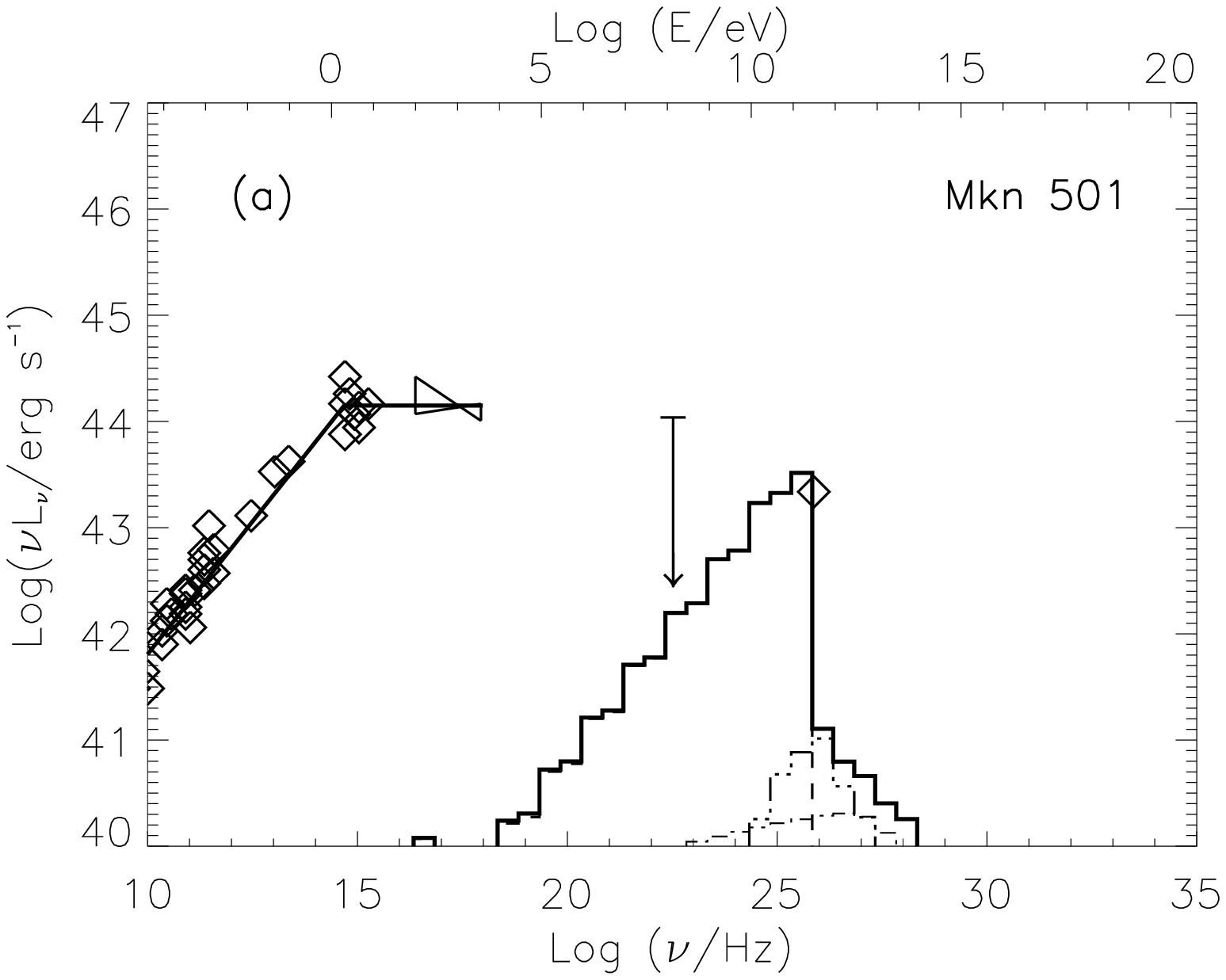, width=7.5cm}
\epsfig{file=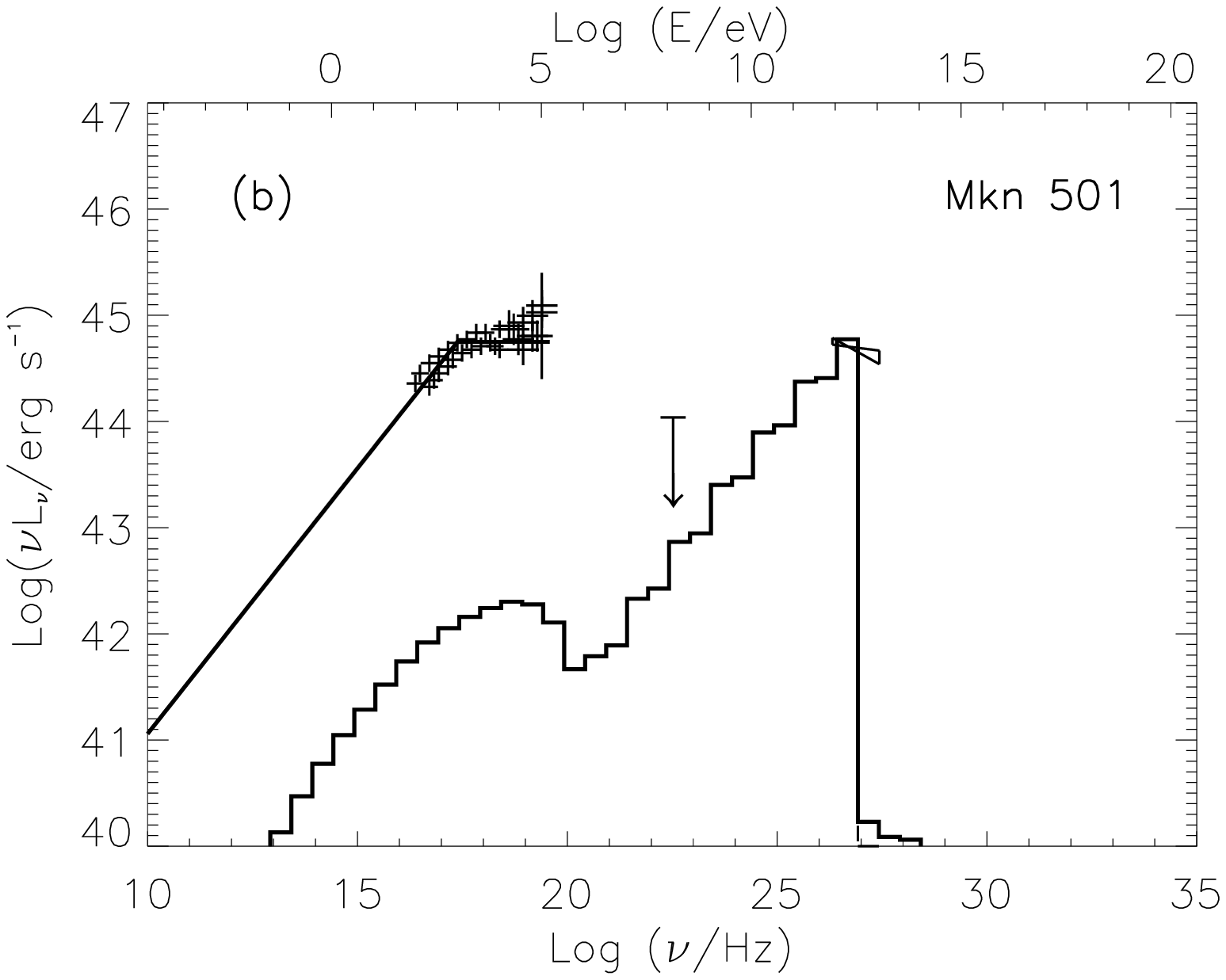, width=7.5cm}
\epsfig{file=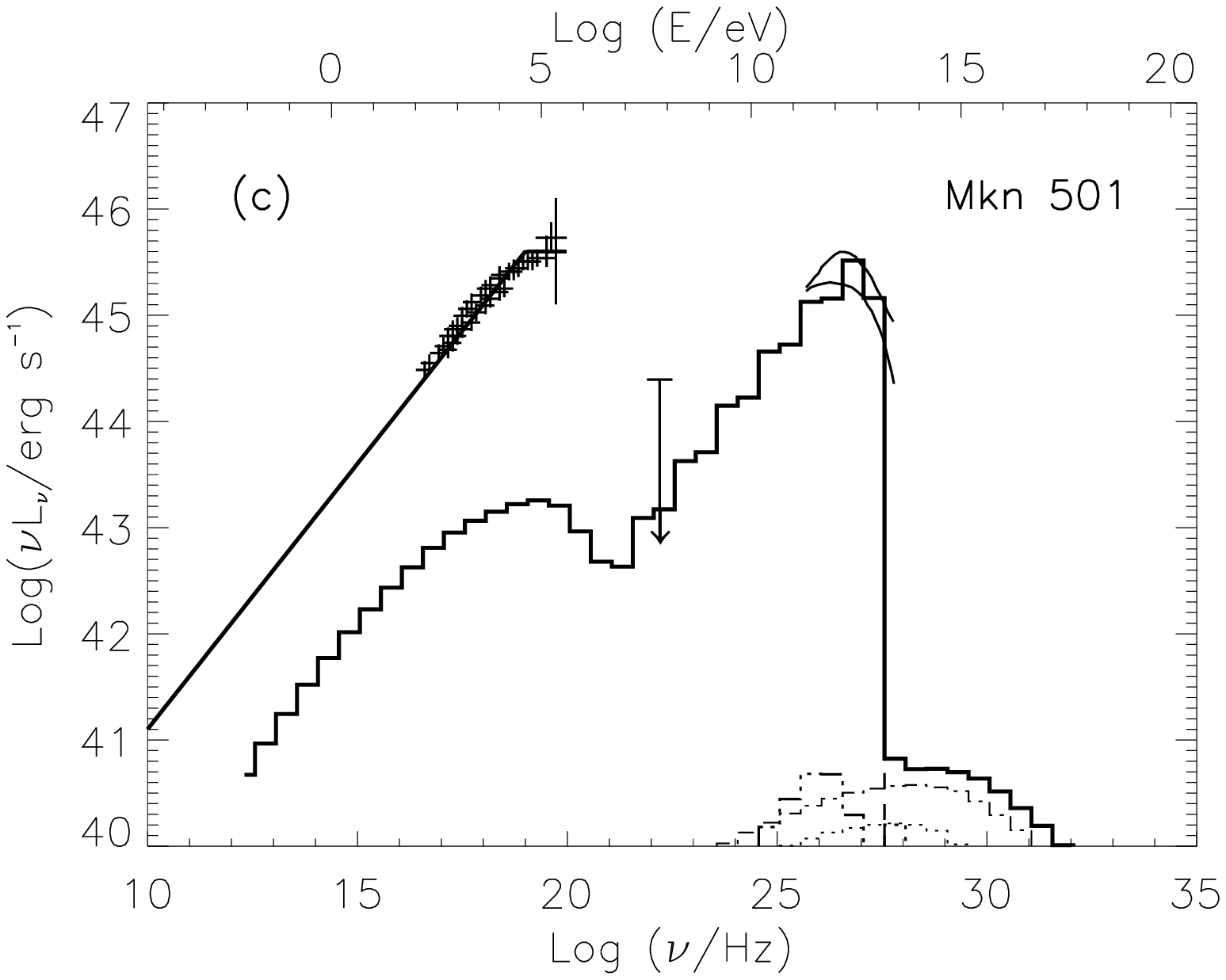, width=7.5cm}
\caption{Modeling the SED of Mkn~501 in different activity states:
(a) quiet state , (b) intermediate state and (c) outburst.
The data are corrected for pair production on the cosmic background
radiation field \cite{BP99}.
Dotted line represents the $\pi^0$-component, dashed-dotted line the
$\pi^{\pm}$-component, dashed-triple dot the $\mu$ synchrotron, dashed
line the $p$ synchrotron component, and the solid line is the sum of all
cascade components. Model parameters are:
a) $B'\approx 20$ G, D = 9,
$R'\approx 5\times 10^{15}$cm, ${u'}_{\rm{phot}} = 7\times 10^{9}$ eV cm$^{-3}$,
${u'}_P = 0.8$ erg/cm$^3$, $\gamma'_{\rm{P,max}} =  10^{10}$, e/p=1.3,
L$_{\rm{jet}}\approx 1.4\times 10^{45}$erg/s, $\eta = 0.05$.
b) $B'\approx 20$ G, D = 11,
$R'\approx 5\times 10^{15}$cm, ${u'}_{\rm{phot}} = 10^{10}$ eV cm$^{-3}$,
${u'}_P = 1.7$ erg/cm$^3$, $\gamma'_{\rm{P,max}} =  4\times 10^{10}$, e/p=0.9,
L$_{\rm{jet}} \approx 1.6\times 10^{45}$erg/s, $\eta = 1$.
c) $B'\approx 20$ G, D = 15,
$R'\approx 5\times 10^{15}$cm, ${u'}_{\rm{phot}} = 3\times 10^{10}$ eV cm$^{-3}$,
${u'}_P = 1.9$ erg/cm$^3$, $\gamma'_{\rm{P,max}} =  4\times 10^{10}$, e/p=0.7,
L$_{\rm{jet}} \approx 4\times 10^{45}$erg/s, $\eta = 1$. }
\label{fig:mrk501flare}
\end{figure}

\begin{figure}[hbt]
\epsfig{file=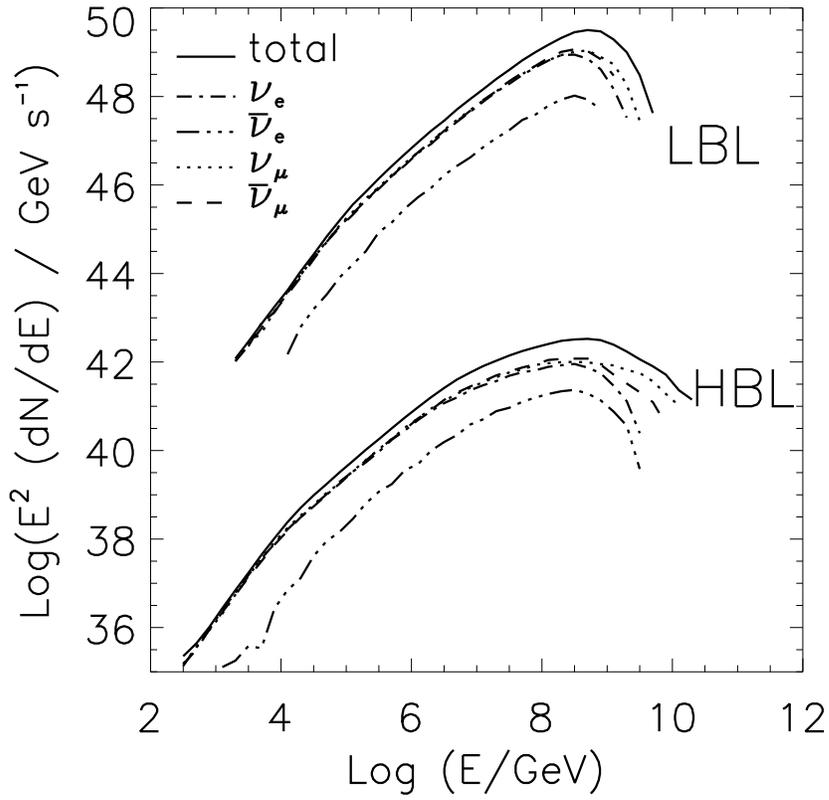}
\caption{Predicted neutrino output for Mkn~421  (labeled ``HBL'') and
PKS~0716+714 (labeled ``LBL'') as modeled in Fig~6.
Antineutrinos from neutron decay are not considered.}
\label{fig:neutrino}
\end{figure}

\begin{figure}[hbt]
\epsfig{file=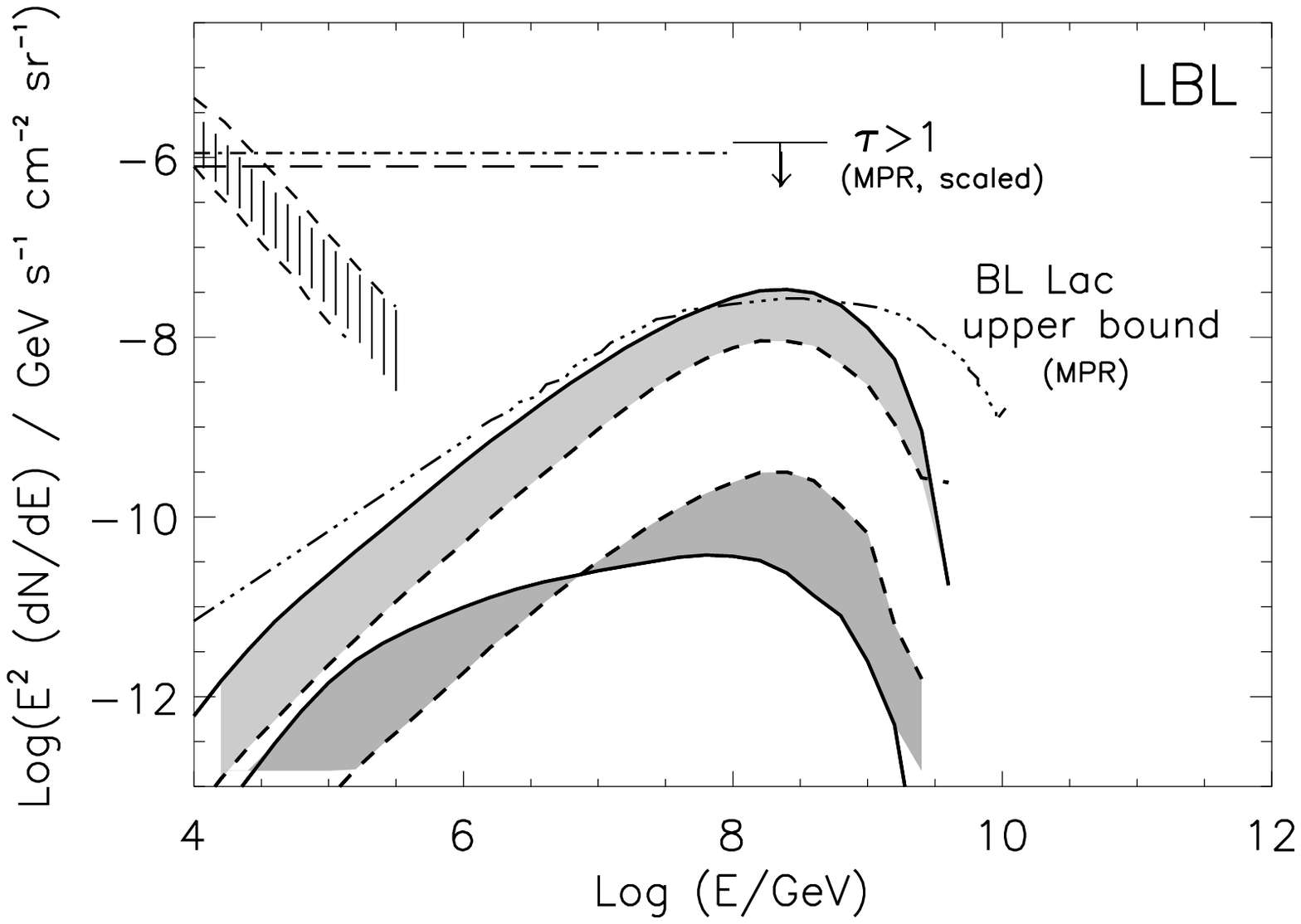, width=7.5cm, height=6cm}
\epsfig{file=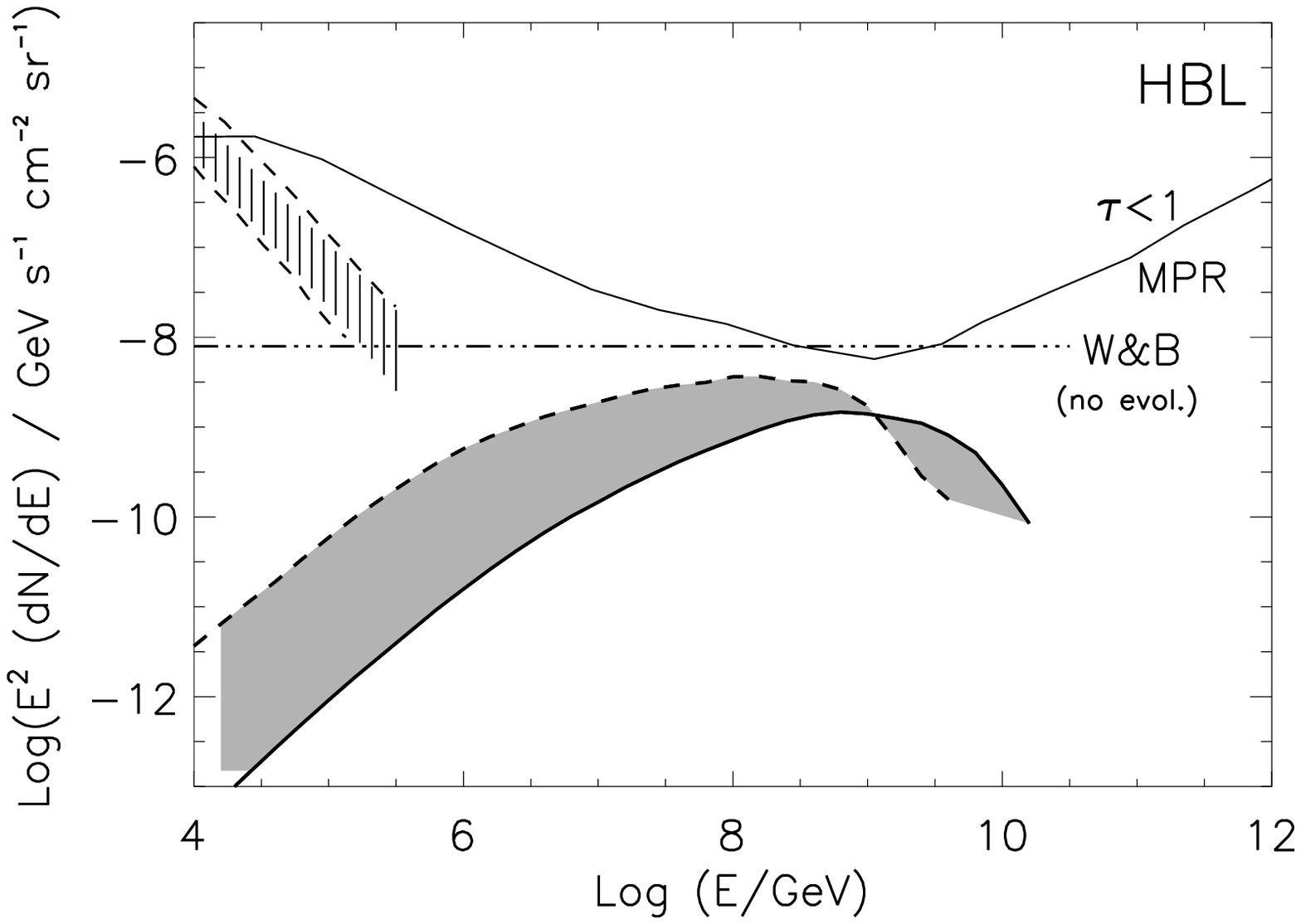, width=7.5cm, height=6cm}
\caption{(a)
Predicted diffuse $\nu_\mu+\bar\nu_\mu$ neutrino spectrum due to
LBLs using the predicted BL~Lac luminosity function of
\cite{UPS91}. The PKS~0716+714/PKS~0537-441 neutrino spectra are
used as a template for a typical LBL, and results in a neutrino
flux estimate corresponding to the upper/lower shaded area,
respectively. The jet frame target photon density
$u'_{\rm{phot}}$ (solid lines) and the ``blob'' radius $R'$ (dashed
lines) are varied to account for the luminosity range in the
luminosity function. Antineutrinos from neutron decay are not considered.
The hatched area on the left represents the
atmospheric neutrino background.  Also shown are blazar
contributions calculated by Stecker et al. \protect \cite{stecker} (chain
line) and Nellen et al. \protect \cite{nellen} (dashed constant line).
The dashed-dotted line corresponds to the upper bound for BL~Lac Objects
derived by \cite{MPR01}, and takes into account both cosmic rays (neutrons)
and gamma-ray emission for these objects.
As can be seen it is in excellent agreement with the predictions in the
present work. The upper limit arrow above the predicted peak power corresponds
to the bolometric bound for sources which are fully opaque ($\tau > 1$) to the
emission of cosmic rays as derived in \cite{MPR01}, taking into account
the bolometric factor of the predicted spectrum. This bound is derived
by considering the observed extragalactic $\gamma$-ray background only.
(b) Predicted diffuse $\nu_\mu+\bar\nu_\mu$ neutrino spectrum due
to HBLs assuming that 10\% of the predicted BL~Lac luminosity
function of \cite{UPS91} is due to HBLs. The Mkn~421 neutrino
spectrum is used as a template for a typical HBL. The jet frame
target photon density $u'_{\rm{phot}}$ (solid lines) and the
``blob'' radius $R'$ (dashed lines) are varied to account for the
luminosity range in the luminosity function. For comparison the
upper bound (thin solid line) for sources that are optically thin to cosmic rays
($\tau<1$) as derived by \cite{MPR01} is shown as well as the upper limit
for sources with no evolution as published in \cite{WB99}.}
\label{fig:diffuse}
\end{figure}

Because $\nu L_{\nu}^{\rm{peak}} \propto R'^2 D^4
u'_{\rm{phot}}$, a range of synchrotron peak luminosities $\nu
L_{\nu}^{\rm{peak}}$ could instead arise due to different sizes
of the emission volume for a constant jet frame target photon
field $u'_{\rm{phot}}$.  To investigate how this would affect the
ENB we have also calculated the contribution of LBLs to the
diffuse neutrino flux in the following way.  From the fits to the
SED of PKS~0716+714 (Fig.~\ref{fig:mrk421pks0716}b) we find a
(jet frame) target photon energy density of $u'_{\rm{phot}}
\approx 10^{11}$ eV cm$^{-3}$, corresponding to $\nu
L_{\nu}^{\rm{peak}} = 10^{47}$ erg/s.  For this $\nu
L_{\nu}^{\rm{peak}}$-value we use the SED of PKS~0716+714
(Fig.~\ref{fig:mrk421pks0716}b) and its associated neutrino
output (Fig.~\ref{fig:neutrino}) is used when calculating the
contribution of LBLs with $\nu L_{\nu}^{\rm{peak}} = 10^{47}$ erg/s to
the ENB.  For other $\nu L_{\nu}^{\rm{peak}}$-values, we assume
the same target photon density $u'_{\rm{phot}} \approx 10^{11}$
eV cm$^{-3}$ but increase or decrease the emission region radius
to give the desired $\nu L_{\nu}^{\rm{peak}}$-value, and
calculate the neutrino output using the SPB model.
The contribution of HBLs to the diffuse neutrino flux is calculated
in a similar way, assuming that the number of HBLs is 10\% of the
number of LBLs \cite{PG95}, but this time using our fits to the
SED of Mkn~421 (Fig.~\ref{fig:mrk421pks0716}a), and its
associated neutrino flux (Fig.~\ref{fig:neutrino}) as a template
for all HBLs with $\nu L_{\nu}^{\rm{peak}} = 8 \times 10^{42}$
erg/s and $u'_{\rm{phot}} \approx 10^{10}$ eV cm$^{-3}$. For
different $L_{\rm 5Ghz}$-values we again increase or decrease the
emission region radius to give the desired $\nu
L_{\nu}^{\rm{peak}}$-value keeping $u'_{\rm{phot}} \approx
10^{10}$ eV cm$^{-3}$.  The upper dashed curves in
Figs.~\ref{fig:diffuse}a,b show the resulting neutrino fluxes. In
reality, the synchrotron luminosity may vary simultaneously due to
both a varying intrinsic synchrotron photon density and a varying
emission volume. The (upper) shaded area therefore gives the
uncertainty in our calculation assuming that the luminosity
conversion ratio $\nu L_{\nu}^{\rm{peak}}/L_{\rm{5GHz}}$ we have
adopted (based on our ``average'' SEDs) and used to convert the
5~GHz luminosity function into a 1.3~eV luminosity function is
correct.

Our estimated diffuse neutrino flux depends strongly on the
$L_{\nu}^{\rm{peak}}/L_{\rm{5GHz}}$-ratio, which in turn is
rather uncertain.  To show this, we have repeated the calculation
for an LBL with a luminosity peak at lower frequency, e.g. PKS
0537-441. Here one finds $\nu L_{\nu}^{\rm{peak}}/\nu
L_{\rm{5GHz}} \approx 60$ for $\nu^{\rm{peak}} \approx 0.07$
eV. The lower shaded area shows the resulting estimates of the
diffuse neutrino flux.

The discussion above shows that there is a large uncertainty in
our predicted diffuse neutrino flux, with $\sim 3$ orders of
magnitude alone due to the uncertainty in the $\nu
L_{\nu}^{\rm{peak}}/\nu L_{\rm{5GHz}}$ ratio, in addition to the
uncertain BL~Lac luminosity function and its LBL fraction.


\section{Summary and Discussion}

We have presented a parameter study of the SPB model proposed
recently to explain the observed spectral energy distribution of
$\gamma$-ray loud BL Lac Objects, i.e.\ HBLs and LBLs.  This model
needs strong magnetic fields together with proton, muon and pion
synchrotron radiation in order to produce the double-humped
structure observed during active phases of $\gamma$-ray emission,
and this is the main difference to the original hadronic ``proton
initiated cascade'' (PIC) model \cite{M93,MB89} which resulted in
a rather featureless $\pi^{0,\pm}$-cascade spectrum.

If LBLs possess denser jet frame synchrotron photon fields than
HBLs, i.e.\ denser target photon fields for $p\gamma$
interactions and cascading in our model, then we have shown that
the high-energy emission in these two types of objects is of
different origin.  While the MeV-TeV radiation from HBLs is
dominated by proton synchrotron radiation, in LBLs there is a
significant contribution from muon synchrotron radiation at
GeV-TeV energies in addition to the proton synchrotron radiation
which dominates at MeV-energies.  This is caused by the
significantly higher pion (and muon) production rate.
Consequently the injected proton spectrum is cut off due to pion
production losses in LBL-like objects, while in HBL-like objects
proton synchrotron radiation is responsible for the cut off in
the proton spectrum. These cutoffs directly translate into the
observed high energy photon cutoff at the source if the Doppler
factor is known.

A further consequence of different intrinsic target photon fields
in HBLs and LBLs is a difference in the ratio of the luminosity
in the high energy hump to low energy hump in their cascade
spectra (due to proton acceleration). Because opacity effects
decisively influence this ratio, the cascade spectra of LBLs have
in general smaller ratios than that of HBLs. Also, the high
energy peak to ``dip'' luminosity appears to be smaller in the
denser LBL-like environments than in HBLs. This is a consequence
of the higher pion production rate in LBLs in comparison to HBLs
which causes the featureless $\pi$-cascade to become important
and fill in the gap between the two humps with proton and muon
radiation.

To demonstrate the difference between LBLs and HBLs in the
SPB-model we have fitted the average observed SED of PKS~0716+714
and Mkn~421. In doing so, we found that HBLs need acceleration
efficiencies of order unity to give high energy hump energies in
the TeV-range, whereas for LBLs acceleration efficiencies of $\sim
10^{-2}$ seem to account for the observations.
LBLs may also produce multi-TeV photons at a lower level than HBLs
despite their higher bolometric luminosity.
The production mechanism is through muon synchrotron radiation
that peaks at a higher energy than the synchrotron radiation of
the primary protons. We therefore suggest that nearby LBLs be
included in the observing source lists of future high-sensitivity
Cherenkov-telescopes (see, e.g., \cite{Krennrich2001}).

Gamma-ray loud BL Lacs are well-known for their flaring activity,
and we have modeled the nearly simultaneous observations of the
intermediate and flaring stage of the famous 1997 giant outburst
of Mkn~501. An increase of the Doppler factor and acceleration
efficiency, together with rising proton and electron density
(leading to a denser intrinsic synchrotron target photon field)
can account for the observations satisfactorily.  Because the observations
were only simultaneous on a one-day time scale at most we believe
that our time-independent code is suitable for this
simulation. At this point we stress the need for time-resolved
simultaneous observations to provide further constrains for blazar
modeling, which then must be carried out using time-dependent
simulation codes.

Although proton and muon synchrotron emission, and their
reprocessed radiation, produce a double-humped structure in
typical blazar jet environments, namely one at X-ray energies,
and another at GeV-TeV energies, the low-energy synchrotron
target photon field dominates over the X-ray hump in the cascade
spectrum in nearly all cases presented here.  This seems to suggest
that the SS-PIC model proposed by \cite{R99,R00}, where the
observed X-ray hump is due to reprocessed proton and muon
synchrotron radiation, is constrained to a rather narrow
parameter range, and this is shown in \cite{R00}. The relatively small
Doppler factors favoured in \cite{R00} imply thick target photon fields,
and consequently significant reprocessing
leading to comparable power at X-ray and sub-TeV energies but making
it difficult to explain the high-energy bump to be at multi-TeV energies.

The dominance of proton synchrotron radiation in HBLs has
recently been used by \cite{aha2000} to consider a blazar model
where all proton synchrotron photons escape the completely
optically thin emission region, and appear as the high-energy
hump in the blazar SED.  This occurs in intrinsically thin or
extremely low energetic ambient photon fields (i.e. very high
Doppler factors $D\approx 10$--$30$ are necessary), where
$p\gamma$-interactions and cascading can be neglected, and
extremely large magnetic fields of $\sim 100$~G are necessary to
fit the observed SED in this case. If the X-ray emission is
produced by synchrotron emission in the strong magnetic field
from secondary electrons which are the result of
$\gamma\gamma$-pair production of the TeV synchrotron photons of
the primary protons with an ambient external infrared photon
field, then the X-ray flare is expected to lag the $\gamma$-ray
flare.  Since acceleration of electrons is faster than of
protons, $t'_{\rm{acc,p}}\gg t'_{\rm{acc,e}}\approx
t'_{\rm{syn,e}}$, typically by a few hours for blazars,
$\gamma$-rays should lag X-rays if the X-rays stem from the
primary co-accelerated electrons.

If electrons are accelerated in the same process as the protons,
$\gamma$-rays from the PIC-processes are in competition with
photons from the leptonic SSC process. SSC-photons contribute
significantly to the escaping radiation if $u'_B \ll
u'_{\rm{phot}}$, i.e.\ $u'_{\rm{phot}} > 10^{13}$eV cm$^{-3}
(B'/30$G$)^2$. So far, all known $\gamma$-ray loud BL Lac objects
have an energy density low-frequency hump $u'_{\rm{phot}} <
10^{12}$eV cm$^{-3}$ for reasonable Doppler factors ($D\approx
10$) and emission volumes, $R'^3\approx (10^{15-17})^3$ cm$^3$.
We conclude that SSC emission is negligible in HBLs and LBLs if
relativistic protons are the main carrier of the dissipated
energy in a highly magnetized jet.

To calculate the total jet luminosity $L_{\rm{jet}}$ measured in
the rest frame of the galaxy we follow the procedure given in
\cite{ProtheroeDonea2002}.  Under the assumption that all
particles (electrons and protons) are relativistic one obtains
\begin{eqnarray}
L_{\rm jet} &=& 4 p'_p \Gamma^2 \beta c \pi R'^2 \left[\chi_p {(\Gamma-1)\over\Gamma} + 1 +
{p'_B\over p'_p}+  {p'_e\over p'_p} \right]\\
&=& {L^{\rm high}_{\rm obs} \over
D^4 \zeta_p} \left[\chi_p {(\Gamma-1)\over\Gamma} + 1 +
{p'_B\over p'_p}+ {\zeta_p L^{\rm low}_{\rm obs} \over \zeta_e L^{\rm
high}_{\rm obs}} \right]
\end{eqnarray}
where $\Gamma = (1 - \beta^2)^{-1} \approx D/2$ is a good
approximation to the Lorentz factor of jets closely aligned to
the line of sight.  $L^{\rm low}_{\rm obs}$ and $L^{\rm
high}_{\rm obs}$ are the observed bolometric luminosities of the
low and high energy component, respectively, and $\zeta_e\approx
1$, $\zeta_p$ are the radiative efficiencies for electrons and
protons. $u'_B=3p'_B$ is the magnetic energy density of a tangled
magnetic field and
\begin{equation}
p'_p = {L^{\rm high}_{\rm obs} \over 4 D^4 \zeta_p \Gamma^2 \beta c
\pi R'^2}
\end{equation}
\begin{equation}
p'_e = {L^{\rm low}_{\rm obs} \over 4 D^4 \zeta_e \Gamma^2 \beta c
\pi R'^2}
\end{equation}
gives the jet-frame pressure of injected relativistic protons and
electrons, respectively, that would apply in the absence of
energy loss mechanisms, and
\begin{eqnarray}
\chi_p &=& {3\over4}\left({p'_e \over p'_p}
{1 \over{\gamma'_e}_1 \ln({\gamma'_e}_2/{\gamma'_e}_1)} + {1 \over {\gamma'_p}_1
\ln({\gamma'_p}_2/{\gamma'_p}_1)}\right)\\
&=& {3\over4}\left({\zeta_p S^{\rm low}_{\rm obs}
\over \zeta_e S^{\rm high}_{\rm obs}}
{1 \over{\gamma'_e}_1 \ln({\gamma'_e}_2/{\gamma'_e}_1)} + {1 \over {\gamma'_p}_1
\ln({\gamma'_p}_2/{\gamma'_p}_1)}\right).
\end{eqnarray}
(Note the erroneous Eq.28+29 in \cite{MP2001a}.)
Here ${\gamma'_e}_1$,${\gamma'_e}_2$ are the lower and upper limit of the injected
electron spectrum, respectively. The ratio of the number of electrons to
protons on injection is then
\begin{equation}
{n'_e \over n'_p} = {p'_e \over p'_p}{m_p \over
m_e}{{\gamma'_p}_1\ln({\gamma'_p}_2/{\gamma'_p}_1) \over{\gamma'_e}_1
\ln({\gamma'_e}_2/{\gamma'_e}_1)}.
\end{equation}
The resulting jet power for HBLs typically lie around
$10^{45}$erg/s, for LBLs typically $10^{47}$erg/s, higher than in
leptonic SSC models but consistent with estimated upper limits
for BL Lacs \cite{CelottiPadovaniGhisellini97}. Injected
$n'_e/n'_p$-ratios are typically $10^{-3}-1$ for HBLs and approximately
unity for LBLs assuming ${\gamma'_e}_1 =2$, i.e. most relativistic
electrons responsible for the low-energy hump in the SED would be
primaries, co-accelerated with the protons.  The addition of cold
electrons possibly needed for charge neutrality in HBLs if $n'_e/n'_p<1$
would add little to the estimated jet power.

A caveat in hadronic models is that most processes are rather
slow in comparison to leptonic interactions.  Indeed, if pion
production dominates the loss processes, variability time scales
below $t_{\rm{var}} \approx 10^3 (10/D) (30\mbox{ G}/B')^2$s at
the highest photon energies would not be expected.  This limit is
based on $u'_B=u'_{\rm{phot}}$ and $\eta=1$, and
assuming that the size of the emission region does not constrain
the variability time scale.  For HBLs proton synchrotron
radiation dominates the loss processes. Hence, the smallest
variability time scale (again provided $R'$ does not determine
$t_{\rm{var}}$) depends on the Doppler factor, magnetic field and
$\eta$, which in turn determines essentially the high
energy photon turnover $\epsilon_{\rm{TeV}}$ (in TeV):
$t_{\rm{var}} \approx 10^4 [(D/10) \epsilon_{\rm{TeV}}
(B'/30\mbox{ G})]^{-0.5}$sec with $\epsilon_{\rm{TeV}} \leq 1.1
(D/10)$ TeV corresponding to $\eta\leq 1$.  Thus, for
extremely high magnetic fields and/or Doppler factors variability
on sub-hour time scales can be reached.

The basic difference between the leptonic SSC model and our presented
hadronic model is the content of the jet: while leptonic models
work with a relativistic electron/positron plasma, our model considers
a relativistic electron/proton jet. For fitting the observed SEDs leptonic
models need significantly smaller magnetic field values
(e.g. $B'=0.497$~G \cite{Gh98} or 0.8~G \cite{pian98} for Mkn~501,
$B'=0.46$~G for PKS~0716+714 and $B'=0.093$~G for Mkn~421 \cite{Gh98})
while the size of the emission region and Doppler factor are
comparable to the values used in this model (e.g. \cite{Gh98} gives
$R'=10^{16}$cm for Mkn~501 and Mkn~421,
and $R'=5\times 10^{16}$cm for PKS~0716+714, $D$=10, 12 and 15 for
Mkn~501, Mkn~421 and PKS~0716+714, respectively, \cite{pian98} gives
$R'=5\times 10^{15}$cm and $D=15$ for the flaring state of Mkn~501).
As a consequence, in leptonic models the particle
energy density is often significantly higher than the magnetic field
energy density.

If the magnetic field component along the
line of sight is much stronger in hadronic models than in leptonic ones,
the rotation measure $RM$ on a length scale of the emission region ($\sim 10^{15}$cm)
may provide a tool to distinguish between the competing models.
Observations of spatial and temporal variability of the $RM$ in
the central parsecs of AGN
suggest that the measured $RM$ is indeed intrinsic to the source and no
foreground effect (e.g. \cite{zavala01}). The observed Faraday rotation
may therefore serve as a probe
of the magnetic field weighted by the electron density in the so-called
Faraday screen along the line of sight on the observed length scale.
In radio galaxies and quasars the Faraday screen is
often considered to be the narrow (NLR) or broad line region (BLR)
(e.g. \cite{taylor}, \cite{zavala01}), and electron densities are derived from
the NLR/BLR optical line strengths.
According to the unified scheme the proposed picture for BL Lac Objects
consists of a relativistic jet that evacuates a cone
through the ionized gas in the nuclear region such that cores of BL Lacs are
not viewed through a dense Faraday screen, and
lower $RM$-values are therefore expected from BL Lac Objects (e.g. \cite{taylor}).
To date, $RM$ measurements from AGN exist only on kpc-pc scales. E.g., for
BL Lacs \cite{zavala02} recently found $RM\sim$ several 100 rad~m$^{-2}$
on 1-50pc scales. Assuming $N_e\approx 10-100$cm$^{-3}$ these values fit to the
strong magnetic fields in hadronic models if the field decays
along the jet as $R^{-1.5\ldots -2}$. High-resolution $RM$ observations
on the central $10^{-3}$~pc scales and a definite identification of the Faraday screen in BL Lacs
are needed to clearly constrain the magnetic field in the gamma ray emission region.

In contrast to leptonic models, models involving pion production
inevitably predict neutrino emission due to the decay of charged
mesons.  In the present work, we predict the
neutrino output of a typical LBL, PKS~0716+714, and a typical
HBL, Mkn~421. If LBLs possess intrinsically denser target photon
fields than HBLs, then within the SPB model, higher meson
production rates are expected in LBLs, leading to a higher
neutrino production rate. The diffuse neutrino background will
therefore be dominated by LBLs, unless HBLs turn out to be
significantly more numerous than LBLs.

To estimate the diffuse neutrino flux we must know the luminosity
functions of HBLs and LBLs, and the neutrino energy spectrum of
HBLs and LBLs as a function of their luminosity.  Because of the
uncertainties in the BL Lac luminosity function and the
conversion from low energy peak to 5 GHz-luminosity, the diffuse
neutrino flux can only be predicted within a large uncertainty of
more than three orders of magnitude. To reduce this uncertainty,
it would be helpful to have a luminosity function for $\nu
L_{\nu}^{\rm{peak}}$ or the bolometric luminosity of the
low-energy hump.

\section*{Acknowledgements}
We thank Alina Donea for helpful discussion. The research of RJP
is supported by the Australian Research Council.
RE and TS are supported in part by NASA Grant NAG5-7009 and by the US
Department of Energy contract DE-FG02 91ER 40626.
AM has been supported by the Quebec Government by a postdoctoral bursary,
and thanks the Bundesministerium f\"ur Bildung und Forschung
for financial support through DESY grant Verbundforschung 05CH1PCA6.
The work of JPR has been supported by the EU-TMR network
Astro-Plasma-Physics (ERBFMRX-CT98-0168).


\appendix

\section{Energy losses from adiabatic jet expansion}

The rate of proton energy change due to adiabatic jet expansion/contraction
is given by
$$
\left(\frac{dE}{dt}\right)'_{\rm{ad}} = -\frac{E'}{3} \vec\bigtriangledown \cdot \vec v
$$
where $\vec v$ is the flow velocity. We consider a velocity field
within the jet which is directed radially outward. On the jet axis at
a distance $z$ from the central engine we can approximate:
$$
\vec\bigtriangledown \cdot \vec v \approx 2 \frac{v}{R'}\frac{dR'}{dz}+\frac{dv}{dz}
$$
with $R'(z)$ the jet radius at distance $z$ in the jet frame. Thus
for a constant speed conical jet $\vec{v} = v\hat{r}$ with opening angle $\Theta$
(i.e. $R' \approx \Theta z$), we find:
$$
\left(\frac{dE}{dt}\right)'_{\rm{ad}} = -\frac{2}{3}\frac{v}{z} E'
$$
with $v=c\sqrt{1-\gamma^{-2}}$ and $dv/dz =c^2/(\gamma^3v)(d\gamma/dz)\approx c/\gamma^3(d\gamma/dz)$
for a relativistic jet. We obtain for the adiabatic loss time scale in the jet frame
$$
t'_{\rm{ad}} = \frac{3}{2} \frac{z}{\gamma c} \approx \frac{R'}{c}.
$$

\newpage

\vspace*{5cm}

\newpage



\vspace*{5cm}

\newpage



\vspace*{5cm}

\begin{thebibliography}{}

\bibitem{DS93} C.D. Dermer \& R. Schlickeiser, {\it ApJ} \ {\bf 416} (1993), 458.

\bibitem{BMS97} M. B\"ottcher, H. Mause \& R. Schlickeiser, {\it A\&A} \ {\bf 324} (1997), 395.

\bibitem{sikora94} M. Sikora, M.C. Begelman \& M.J. Rees, {\it ApJ} \ {\bf 421} (1994), 153.

\bibitem{blaze2000} M. Blazejowski, M. Sikora, R. Moderski \& G.M. Madejski, {\it ApJ} \ {\bf 545} (2000), 107.

\bibitem{DoneaProtheroe2002} A.-C. Donea and R.J. Protheroe, {\it Astropart.Phys.}, submitted

\bibitem{MGC92} L. Maraschi, G. Ghisellini \& A. Celotti, {\it ApJ} \ {\bf 397} (1992), L5.

\bibitem{BM96} S.D. Bloom \& A.P. Marscher, {\it ApJ} \ {\bf 461} (1996), 657.

\bibitem{MB89} K. Mannheim \& P.L. Biermann, {\it A\&A} \ {\bf 221} (1989), 211.

\bibitem{M93} K. Mannheim, {\it A\&A} \ {\bf 269} (1993), 67.

\bibitem{MP2000} A. M\"ucke \& R.J. Protheroe,
{\it Proc. workshop "GeV-TeV Astrophysics: Toward a Major Atmospheric Cherenkov Telescope VI",
AIP Conf. Proc.} \ { Vol 515} (2000), 149, eds.: B.D. Dingus et al.

\bibitem{MP2001a} A. M\"ucke \& R.J. Protheroe, {\it Astropart. Phys}\ {\bf 15} (2001),
121.

\bibitem{RM98} J. Rachen \& P. M\'esz\'aros, {\it Phys. Rev D}\ {\bf 58} (1998), 123005.

\bibitem{P96} R.J. Protheroe, {\it Proc. IAU Colloq. 163}, Accretion Phenomena and
Related Outflows, ed. D. Wickramasinghe et al. (1996).

\bibitem{LM2000} J. Learned \& K. Mannheim, {\it Ann.Rev.Nucl.Part.Sci.} \ {\bf 50} (2000), 679.

\bibitem{PJ96} R.J. Protheroe \& P. Johnson, {\it Astropart.Phys.} \ {\bf 4} (1996), 253,
\& erratum \ {\bf 5} (1996), 215.

\bibitem{SOPHIA} A. M\"ucke et al.,{\it Comm.Phys.Comp.}\ {\bf 124} (2000), 290.

\bibitem{Protheroe_variabilitysize} R.J. Protheroe, {\it PASA}, submitted.

\bibitem{schlicki} M. Pohl \& R. Schlickeiser, {\it A\&A} \ {\bf 354} (2000), 395.

\bibitem{BZ77} R.D. Blandford \& R.L. Znajek, {\it MNRAS}\ {\bf 179} (1977), 433.

\bibitem{PG95} P. Padovani \& P. Giommi, {\it ApJ}\ {\bf 444} (1995), 567 (PG95).

\bibitem{perl98} E. Perlman, P. Padovani, P. Giommi, R. Sambruna, L.R. Jones,
A. Tzioumis, J. Reynolds, {\it AJ}\ {\bf 115} (1998), 1253.

\bibitem{LM99} S.A. Laurant-Muehleisen,
R.I. Kollgaard, E.D. Feigelson, W. Brinkmann, J. Siebert,
1998, {\it ApJS}\ {\bf 525} (1999), 127.

\bibitem{cac99} A. Caccianiga, T. Maccacaro, A. Wolter, R. Della Ceca, I.M. Gioia,
{\it ApJ}\ {\bf 513} (1999), 51.

\bibitem{Gh98} G. Ghisellini et al., {\it MNRAS}\ {\bf 301} (1998), 451.

\bibitem{R99} J.P. Rachen 1999,
in: Proc. of ``GeV-TeV Astrophysics: Toward a Major Atmospheric Cherenkov
Telescope V'', Snowbird, Utah (August, 1999).

\bibitem{aha2000} F.A. Aharonian, {\it New Astron.}\ {\bf 5} (2000), 377.

\bibitem{BP99} W. Bednarek \& R.J. Protheroe, {\it MNRAS}\ {\bf 310} (1999), 577.

\bibitem{henri99} G. Henri, G. Pelletier, P.O. Petrucci, N. Renaud,
{\it Astropart. Phys.}\ {\bf 11} (1999), 347.


\bibitem{haswell92} C.A. Haswell, T. Tajima, J.J. Sakai, {\it ApJ}\
{\bf 401} (1992), 495.

\bibitem{GLAST} N. Gehrels \& P. Michelson, {\it Astropart. Phys.} \ {\bf 11} (1999), 277.

\bibitem{pian98} E. Pian et al., {\it ApJ} \ {\bf 492} (1998), L17.

\bibitem{MP2001b} A. M\"ucke \& R.J. Protheroe, Proc. 27th Int. Cosmic Ray Conf., Hamburg/Germany,
{\bf Vol. 3}, (2001), 1153.

\bibitem{sree98} P. Sreekumar et al., {\it ApJ} \ {\bf 494} (1998), 523.

\bibitem{CM98} J. Chiang \& R. Mukherjee, {\it ApJ} \ {\bf 496} (1998), 752.

\bibitem{WB99} E. Waxman \& J. Bahcall, {\it Phys. Rev. D} \ {\bf 59} (1999), 023002.

\bibitem{MPR01} K. Mannheim, R.J. Protheroe \& J.P. Rachen, {\it Phys Rev D} \ {\bf 63} (2001), 023003.

\bibitem{pado01} P. Padovani, {\it Proc. workshop "Blazar Demographics and
Physics", ASP Conf. Series} \ {\bf 227} (2001),  eds.: P. Padovani \& C.M. Urry.

\bibitem{giom01} P. Giommi \& A. Pellizzoni, {\it Proc. workshop "Blazar Demographics and
Physics", ASP Conf. Series} \ {\bf 227} (2001),  eds.: P. Padovani \& C.M. Urry.

\bibitem{Cacc01} A. Caccianiga, T. Maccacaro, A. Wolter, R. Della Ceca, I.M. Gioia,
{\it ApJ} \ {\bf 566} (2002), 181.

\bibitem{UPS91} C.M. Urry, P. Padovani \& M. Stickel, {\it ApJ} \ {\bf 382} (1991), 501.

\bibitem{Krennrich2001} F. Krennrich, in: ``TeV gamma-ray astronomy in the New Millennium'',
7th Taipei Astrophysics Workshop on Cosmic Rays in the Universe, Ed.~C.-M. Ko, ASP Conference Series, Vol.~241 (2001), 141.

\bibitem{R00} J. Rachen, in: {\it ``Gamma-2000''}, Heidelberg,
AIP Proceedings, eds.: F.A. Aharonian \& H.J. V\"olk, {\bf 558} (2000), 704.

\bibitem{ProtheroeDonea2002} R.J. Protheroe and A.-C. Donea, to be submitted.

\bibitem{CelottiPadovaniGhisellini97} A. Celotti, P. Padovani, G. Ghisellini, {\it MNRAS}\ {\bf 286} (1997), 415.

\bibitem{kerr95} A.D. Kerrick et al., {\it ApJ} \ {\bf 452} (1995), 588.

\bibitem{stecker} F.W. Stecker \& M.H. Salamon, {\it Space Sci.Rev.} \ {\bf 75} (1996), 341.

\bibitem{nellen} L. Nellen, K. Mannheim \& P.L. Biermann, {\it Phys.Rev.D}\ {\bf 47} (1993), 5270.

\bibitem{zavala01} R.T. Zavala \& G.B. Taylor, {\it American Astronomical Society Meeting} \ {\bf 199} (2001).

\bibitem{taylor} G.B. Taylor, {\it ApJ} \ {\bf 533} (2000), 95.

\bibitem{zavala02} R.T. Zavala \& G.B. Taylor, {\it American Astronomical Society Meeting} \ {\bf 200} (2002).


\end{thebibliography}
\end{document}